\documentclass[a4paper,onecolumn,superscriptaddress,11pt,accepted=2022-08-03]{quantumarticle}
\pdfoutput=1
\usepackage{pifont}
\newcommand{\cmark}{\ding{51}}
\newcommand{\xmark}{\ding{55}}
\usepackage{makecell}
\usepackage{qcircuit}
\usepackage{cancel}
\usepackage{soul}
\usepackage{changes,subcaption}
\usepackage{hyperref}

\usepackage{url}
\definecolor{darkblue}{RGB}{0,0,149}
\definecolor{darkgreen}{rgb}{0.0, 0.5, 0.13}

\def\Hy{{\sf H}}
\def\A{{\sf A}}
\def\B{{\sf B}}

\def\G{{\sf G}}

\def\Q{{\sf Q}}

\def\R{{\mathbb R}}

\usepackage{rotating}
\usepackage{floatpag}
\rotfloatpagestyle{empty}
\renewcommand{\theenumi}{\arabic{enumi}}
\usepackage{hyperref}
\usepackage{dsfont}
\usepackage{amsmath,amsthm,amssymb}
\usepackage{xspace,enumerate,epsfig}%color,
\usepackage{graphicx}
\graphicspath{{.}{./figures/}}
\usepackage{marvosym}

\usepackage{tikzfig}
\usepackage{stmaryrd}
\usepackage{docmute}
\usepackage{keycommand}

\newcommand{\C}{\ensuremath{\mathbb{C}}}

\newkeycommand{\pointmap}[style=point][1]{\,\tikz{\node[style=\commandkey{style}] (x) at (0,-0.3) {$#1$};\node [style=none] (3) at (0, 1.0) {};\node [style=none] (3) at (0, -1.0) {};\draw (x)--(0,0.7);}\,}

\newkeycommand{\typedkpoint}[style=kpoint,edgestyle=][2]{%
\,\begin{tikzpicture}
  \begin{pgfonlayer}{nodelayer}
    \node [style=none] (0) at (0, 1) {};
    \node [style=\commandkey{style}] (1) at (0, -0.75) {$#2$};
    \node [style=right label] (2) at (0.25, 0.5) {$#1$};
  \end{pgfonlayer}
  \begin{pgfonlayer}{edgelayer}
    \draw [\commandkey{edgestyle}] (1) to (0.center);
  \end{pgfonlayer}
\end{tikzpicture}\,}

\newkeycommand{\typedkpointdag}[style=kpoint adjoint,edgestyle=][2]{%
\,\begin{tikzpicture}
  \begin{pgfonlayer}{nodelayer}
    \node [style=none] (0) at (0, -1) {};
    \node [style=\commandkey{style}] (1) at (0, 0.75) {$#2$};
    \node [style=right label] (2) at (0.25, -0.5) {$#1$};
  \end{pgfonlayer}
  \begin{pgfonlayer}{edgelayer}
    \draw [\commandkey{edgestyle}] (0.center) to (1);
  \end{pgfonlayer}
\end{tikzpicture}\,}

\newcommand{\bistateadj}[1]{\,\begin{tikzpicture}[yshift=-3mm]
  \begin{pgfonlayer}{nodelayer}
    \node [style=kpointadj, minimum width=1 cm, inner sep=2pt] (0) at (0, 0.5) {$\psi$};
    \node [style=none] (1) at (-0.75, 0.25) {};
    \node [style=none] (2) at (0.75, 0.25) {};
    \node [style=none] (3) at (-0.75, -0.5) {};
    \node [style=none] (4) at (0.75, -0.5) {};
  \end{pgfonlayer}
  \begin{pgfonlayer}{edgelayer}
    \draw (1.center) to (3.center);
    \draw (2.center) to (4.center);
  \end{pgfonlayer}
\end{tikzpicture}\,}

\newkeycommand{\pointketbra}[style1=copoint,style2=point][2]{\,%
\begin{tikzpicture}
  \begin{pgfonlayer}{nodelayer}
    \node [style=none] (0) at (0, -1.5) {};
    \node [style=\commandkey{style1}] (1) at (0, -0.75) {$#1$};
    \node [style=\commandkey{style2}] (2) at (0, 0.75) {$#2$};
    \node [style=none] (3) at (0, 1.5) {};
  \end{pgfonlayer}
  \begin{pgfonlayer}{edgelayer}
    \draw (0.center) to (1);
    \draw (2) to (3.center);
  \end{pgfonlayer}
\end{tikzpicture}\,}

\newkeycommand{\twocopointketbra}[style1=copoint,style2=copoint,style3=point][3]{\,%
\begin{tikzpicture}
  \begin{pgfonlayer}{nodelayer}
    \node [style=none] (0) at (-0.75, -1.5) {};
    \node [style=none] (1) at (0.75, -1.5) {};
    \node [style=\commandkey{style1}] (2) at (-0.75, -0.75) {$#1$};
    \node [style=\commandkey{style2}] (3) at (0.75, -0.75) {$#2$};
    \node [style=\commandkey{style3}] (4) at (0, 0.75) {$#3$};
    \node [style=none] (5) at (0, 1.5) {};
  \end{pgfonlayer}
  \begin{pgfonlayer}{edgelayer}
    \draw (0.center) to (2);
    \draw (1.center) to (3);
    \draw (4) to (5.center);
  \end{pgfonlayer}
\end{tikzpicture}\,}

\newkeycommand{\twopointketbra}[style1=copoint,style2=point,style3=point][3]{\,%
\begin{tikzpicture}
  \begin{pgfonlayer}{nodelayer}
    \node [style=none] (0) at (0, -1.5) {};
    \node [style=none] (1) at (-0.75, 1.5) {};
    \node [style=\commandkey{style1}] (2) at (0, -0.75) {$#1$};
    \node [style=\commandkey{style2}] (3) at (-0.75, 0.75) {$#2$};
    \node [style=\commandkey{style3}] (4) at (0.75, 0.75) {$#3$};
    \node [style=none] (5) at (0.75, 1.5) {};
  \end{pgfonlayer}
  \begin{pgfonlayer}{edgelayer}
    \draw (0.center) to (2);
    \draw (1.center) to (3);
    \draw (4) to (5.center);
  \end{pgfonlayer}
\end{tikzpicture}\,}

\newkeycommand{\pointbraket}[style1=point,style2=copoint][2]{\,%
\begin{tikzpicture}
  \begin{pgfonlayer}{nodelayer}
    \node [style=\commandkey{style1}] (0) at (0, -0.5) {$#1$};
    \node [style=\commandkey{style2}] (1) at (0, 0.5) {$#2$};
  \end{pgfonlayer}
  \begin{pgfonlayer}{edgelayer}
    \draw (0) to (1);
  \end{pgfonlayer}
\end{tikzpicture}\,}

\newkeycommand{\kpointbraket}[style1=kpoint,style2=kpoint adjoint][2]{\,%
\begin{tikzpicture}
  \begin{pgfonlayer}{nodelayer}
    \node [style=\commandkey{style1}] (0) at (0, -0.75) {$#1$};
    \node [style=\commandkey{style2}] (1) at (0, 0.75) {$#2$};
  \end{pgfonlayer}
  \begin{pgfonlayer}{edgelayer}
    \draw (0) to (1);
  \end{pgfonlayer}
\end{tikzpicture}\,}

\newkeycommand{\kpointmap}[style1=kpoint,style2=map][2]{\,%
\begin{tikzpicture}
  \begin{pgfonlayer}{nodelayer}
    \node [style=\commandkey{style1}] (0) at (0, -1.1) {$#1$};
    \node [style=\commandkey{style2}] (1) at (0, 0.2) {$#2$};
    \node [style=none] (2) at (0, 1.5) {};
  \end{pgfonlayer}
  \begin{pgfonlayer}{edgelayer}
    \draw (0) to (1);
    \draw (1) to (2.center);
  \end{pgfonlayer}
\end{tikzpicture}%
\,}

\newkeycommand{\sandwichtwo}[style1=point,style2=map,style3=map,style4=copoint][4]{\,%
\begin{tikzpicture}
  \begin{pgfonlayer}{nodelayer}
    \node [style=\commandkey{style2}] (0) at (0, -0.75) {$#2$};
    \node [style=\commandkey{style3}] (1) at (0, 0.75) {$#3$};
    \node [style=\commandkey{style1}] (2) at (0, -2) {$#1$};
    \node [style=\commandkey{style4}] (3) at (0, 2) {$#4$};
  \end{pgfonlayer}
  \begin{pgfonlayer}{edgelayer}
    \draw (2) to (0);
    \draw (0) to (1);
    \draw (1) to (3);
  \end{pgfonlayer}
\end{tikzpicture}\,}

\newkeycommand{\longbraket}[style1=point,style2=copoint][2]{\,%
\begin{tikzpicture}
  \begin{pgfonlayer}{nodelayer}
    \node [style=\commandkey{style1}] (0) at (0, -2) {$#1$};
    \node [style=\commandkey{style2}] (1) at (0, 2) {$#2$};
  \end{pgfonlayer}
  \begin{pgfonlayer}{edgelayer}
    \draw (0) to (1);
  \end{pgfonlayer}
\end{tikzpicture}\,}

\newkeycommand{\onb}[style=point]{\ensuremath{\left\{\tikz{\node[style=\commandkey{style}] (x) at (0,-0.3) {$j$};\draw (x)--(0,0.7);}\right\}}\xspace}

\newkeycommand{\copointmap}[style=copoint][1]{\,\tikz{\node[style=\commandkey{style}] (x) at (0,0.3) {$#1$};\draw (x)--(0,-0.7);}\,}

\newcommand{\ket}[1]{\ensuremath{\left|  #1 \right\rangle}}

\tikzstyle{cdiag}=[matrix of math nodes, row sep=3em, column sep=3em, text height=1.5ex, text depth=0.25ex,inner sep=0.5em]
\tikzstyle{arrow above}=[transform canvas={yshift=0.5ex}]
\tikzstyle{arrow below}=[transform canvas={yshift=-0.5ex}]

\usetikzlibrary{shapes.multipart}

\tikzstyle{bwSpider}=[
       rectangle split,
       rectangle split parts=2,
       rectangle split part fill={black,white},
 minimum size=3.6 mm, inner sep=-2mm, draw=black,scale=0.5,rounded corners=0.8 mm
       ]
 \tikzstyle{wbSpider}=[
       rectangle split,
       rectangle split parts=2,
       rectangle split part fill={white,black},
 minimum size=3.6 mm, inner sep=-2mm, draw=black,scale=0.5,rounded corners=0.8 mm
       ]
\tikzstyle{qWire}=[line width = 1pt, color=black]
\tikzstyle{cWire}=[color=gray,thin]%[densely dotted, thick]
\tikzstyle{env}=[copoint,regular polygon rotate=0,minimum width=0.2cm, fill=black]

\tikzstyle{probs}=[shape=semicircle,fill=white,draw=black,shape border rotate=180,minimum width=1.2cm]

\tikzstyle{every picture}=[baseline=-0.25em,scale=0.5]
\tikzstyle{dotpic}=[] % for backwards-compatibility
\tikzstyle{diredges}=[every to/.style={diredge}]
\tikzstyle{math matrix}=[matrix of math nodes,left delimiter=(,right delimiter=),inner sep=2pt,column sep=1em,row sep=0.5em,nodes={inner sep=0pt},text height=1.5ex, text depth=0.25ex]

\tikzstyle{inline text}=[text height=1.5ex, text depth=0.25ex,yshift=0.5mm]
\tikzstyle{label}=[font=\footnotesize,text height=1.5ex, text depth=0.25ex,yshift=0.5mm]
\tikzstyle{left label}=[label,anchor=east,xshift=1.5mm]
\tikzstyle{right label}=[label,anchor=west,xshift=-1mm]

\tikzstyle{braceedge}=[decorate,decoration={brace,amplitude=2mm,raise=-1mm}]
\tikzstyle{small braceedge}=[decorate,decoration={brace,amplitude=1mm,raise=-1mm}]

\tikzstyle{doubled}=[line width=1.6pt] % set the line width for all doubled (quantum) maps/wires
\tikzstyle{boldedge}=[doubled,shorten <=-0.17mm,shorten >=-0.17mm]
\tikzstyle{boldedgegray}=[doubled,gray,shorten <=-0.17mm,shorten >=-0.17mm]
\tikzstyle{singleedgegray}=[gray]%,shorten <=-0.1mm,shorten >=-0.1mm]

\tikzstyle{semidoubled}=[line width=1.4pt] % set the line width for all doubled (quantum) maps/wires
\tikzstyle{semiboldedgegray}=[semidoubled,gray,shorten <=-0.17mm,shorten >=-0.17mm]

\tikzstyle{boxedge}=[semiboldedgegray]

\tikzstyle{boldedgedashed}=[very thick,dashed,shorten <=-0.17mm,shorten >=-0.17mm]
\tikzstyle{vboldedgedashed}=[doubled,dashed,shorten <=-0.17mm,shorten >=-0.17mm]
\tikzstyle{left hook arrow}=[left hook-latex]
\tikzstyle{right hook arrow}=[right hook-latex]
\tikzstyle{sembracket}=[line width=0.5pt,shorten <=-0.07mm,shorten >=-0.07mm]

\tikzstyle{causal edge}=[->,thick,gray]
\tikzstyle{causal nondir}=[thick,gray]
\tikzstyle{timeline}=[thick,gray, dashed]

\tikzstyle{cedge}=[<->,thick,gray!70!white]

\tikzstyle{empty diagram}=[draw=gray!40!white,dashed,shape=rectangle,minimum width=1cm,minimum height=1cm]
\tikzstyle{empty diagram small}=[draw=gray!50!white,dashed,shape=rectangle,minimum width=0.6cm,minimum height=0.5cm]

\tikzstyle{dot}=[inner sep=0mm,minimum width=2mm,minimum height=2mm,draw,shape=circle]
\tikzstyle{leak}=[white dot, shape=regular polygon, minimum size=3.3 mm, regular polygon sides=3, outer sep=-0.2mm, regular polygon rotate=270]
\tikzstyle{proj}=[draw,fill=white,chamfered rectangle,chamfered rectangle angle=30, minimum width=2mm, minimum height= 1mm,scale=0.5, outer sep=-0.2mm]
\tikzstyle{wide proj}=[draw,fill=white,chamfered rectangle,chamfered rectangle angle=30, minimum width=15mm,minimum height=1mm,scale=0.5, outer sep=-0.2mm]
\tikzstyle{very wide proj}=[draw,fill=white,chamfered rectangle,chamfered rectangle angle=30, minimum width=25mm,minimum height=1mm,scale=0.5, outer sep=-0.2mm]
\tikzstyle{very very wide proj}=[draw,fill=white,chamfered rectangle,chamfered rectangle angle=30, minimum width=35mm,minimum height=1mm,scale=0.5, outer sep=-0.2mm]
\tikzstyle{preleak}=[proj]

\tikzstyle{split proj out}=[regular polygon,regular polygon sides=3,draw,scale=0.75,inner sep=-0.5pt,minimum width=3.3mm,fill=white,regular polygon rotate=180]
\tikzstyle{split proj in}=[regular polygon,regular polygon sides=3,draw,scale=0.75,inner sep=-0.5pt,minimum width=3.3mm,fill=white]
\tikzstyle{Vleak}=[white dot, shape=regular polygon, minimum size=3.3 mm, regular polygon sides=3, outer sep=-0.2mm, regular polygon rotate=90]
\tikzstyle{dleak}=[white dot, line width=1.6pt, shape=regular polygon, minimum size=3.3 mm, regular polygon sides=3, outer sep=-0.2mm, regular polygon rotate=270]

\tikzstyle{Wsquare}=[white dot, shape=regular polygon, rounded corners=0.8 mm, minimum size=3.3 mm, regular polygon sides=3, outer sep=-0.2mm]
\tikzstyle{Wsquareadj}=[white dot, shape=regular polygon, rounded corners=0.8 mm, minimum size=3.3 mm, regular polygon sides=3, outer sep=-0.2mm, regular polygon rotate=180]
\tikzstyle{ddot}=[inner sep=0mm, doubled, minimum width=2.5mm,minimum height=2.5mm,draw,shape=circle]

\tikzstyle{black dot}=[dot,fill=black]
\tikzstyle{white dot}=[dot,fill=white,,text depth=-0.2mm]
\tikzstyle{white Wsquare}=[Wsquare,fill=gray,,text depth=-0.2mm]
\tikzstyle{white Wsquareadj}=[Wsquareadj,fill=white,,text depth=-0.2mm]
\tikzstyle{green dot}=[white dot] % for backwards-compatibility
\tikzstyle{gray dot}=[dot,fill=gray!40!white,,text depth=-0.2mm]
\tikzstyle{red dot}=[gray dot] % for backwards-compatibility

\tikzstyle{black ddot}=[ddot,fill=black]
\tikzstyle{white ddot}=[ddot,fill=white]
\tikzstyle{gray ddot}=[ddot,fill=gray!40!white]

\tikzstyle{gray edge}=[gray!60!white]

\tikzstyle{small dot}=[inner sep=0.5mm,minimum width=0pt,minimum height=0pt,draw,shape=circle]

\tikzstyle{small black dot}=[small dot,fill=black]
\tikzstyle{small white dot}=[small dot,fill=white]
\tikzstyle{small gray dot}=[small dot,fill=gray!40!white]

\tikzstyle{causal dot}=[inner sep=0.4mm,minimum width=0pt,minimum height=0pt,draw=white,shape=circle,fill=gray!40!white]

\tikzstyle{phase dimensions}=[minimum size=5mm,font=\footnotesize,rectangle,rounded corners=2.5mm,inner sep=0.2mm,outer sep=-2mm]
\tikzstyle{dphase dimensions}=[minimum size=5mm,font=\footnotesize,rectangle,rounded corners=2.5mm,inner sep=0.2mm,outer sep=-2mm]
\tikzstyle{white phase dot}=[dot,fill=white,phase dimensions]
\tikzstyle{white phase ddot}=[ddot,fill=white,dphase dimensions]

\tikzstyle{white rect ddot}=[draw=black,fill=white,doubled,minimum size=5mm,font=\footnotesize,rectangle,rounded corners=2.5mm,inner sep=0.2mm]
\tikzstyle{gray rect ddot}=[draw=black,fill=gray!40!white,doubled,minimum size=6mm,font=\footnotesize,rectangle,rounded corners=3mm]

\tikzstyle{gray phase dot}=[dot,fill=gray!40!white,phase dimensions]
\tikzstyle{gray phase ddot}=[ddot,fill=gray!40!white,dphase dimensions]
\tikzstyle{grey phase dot}=[gray phase dot]
\tikzstyle{grey phase ddot}=[gray phase ddot]

\tikzstyle{small phase dimensions}=[minimum size=4mm,font=\tiny,rectangle,rounded corners=2mm,inner sep=0.2mm,outer sep=-2mm]
\tikzstyle{small dphase dimensions}=[minimum size=4mm,font=\tiny,rectangle,rounded corners=2mm,inner sep=0.2mm,outer sep=-2mm]

\tikzstyle{small gray phase dot}=[dot,fill=gray!40!white,small phase dimensions]
\tikzstyle{small gray phase ddot}=[ddot,fill=gray!40!white,small dphase dimensions]

\tikzstyle{small map}=[draw,shape=rectangle,minimum height=4mm,minimum width=4mm,fill=white]

\tikzstyle{cnot}=[fill=white,shape=circle,inner sep=-1.4pt]

\tikzstyle{asym hadamard}=[fill=white,draw,shape=NEbox,inner sep=0.6mm,font=\footnotesize,minimum height=4mm]
\tikzstyle{asym hadamard conj}=[fill=white,draw,shape=NWbox,inner sep=0.6mm,font=\footnotesize,minimum height=4mm]
\tikzstyle{asym hadamard dag}=[fill=white,draw,shape=SEbox,inner sep=0.6mm,font=\footnotesize,minimum height=4mm]

\tikzstyle{hadamard}=[fill=white,draw,inner sep=0.6mm,font=\footnotesize,minimum height=4mm,minimum width=4mm]
\tikzstyle{small hadamard}=[fill=white,draw,inner sep=0.6mm,minimum height=1.5mm,minimum width=1.5mm]
\tikzstyle{small hadamard rotate}=[small hadamard,rotate=45]
\tikzstyle{dhadamard}=[hadamard,doubled]
\tikzstyle{small dhadamard}=[small hadamard,doubled]
\tikzstyle{small dhadamard rotate}=[small hadamard rotate,doubled]
\tikzstyle{antipode}=[white dot,inner sep=0.3mm,font=\footnotesize]

\tikzstyle{scalar}=[diamond,draw,inner sep=0.5pt,font=\small]
\tikzstyle{dscalar}=[diamond,doubled, draw,inner sep=0.5pt,font=\small]

\tikzstyle{small box}=[rectangle,inline text,fill=white,draw,minimum height=5mm,yshift=-0.5mm,minimum width=5mm,font=\small]
\tikzstyle{small gray box}=[small box,fill=gray!30]
\tikzstyle{medium box}=[rectangle,inline text,fill=white,draw,minimum height=5mm,yshift=-0.5mm,minimum width=10mm,font=\small]
\tikzstyle{square box}=[small box] % for backwards-compatibility
\tikzstyle{medium gray box}=[small box,fill=gray!30]
\tikzstyle{semilarge box}=[rectangle,inline text,fill=white,draw,minimum height=5mm,yshift=-0.5mm,minimum width=12.5mm,font=\small]
\tikzstyle{large box}=[rectangle,inline text,fill=white,draw,minimum height=5mm,yshift=-0.5mm,minimum width=15mm,font=\small]
\tikzstyle{large gray box}=[small box,fill=gray!30]

\tikzstyle{Bayes box}=[rectangle,fill=black,draw, minimum height=3mm, minimum width=3mm]

\tikzstyle{gray square point}=[small box,fill=gray!50]

\tikzstyle{dphase box white}=[dhadamard]
\tikzstyle{dphase box gray}=[dhadamard,fill=gray!50!white]
\tikzstyle{phase box white}=[hadamard]
\tikzstyle{phase box gray}=[hadamard,fill=gray!50!white]

\tikzstyle{point}=[regular polygon,regular polygon sides=3,draw,scale=0.75,inner sep=-0.5pt,minimum width=9mm,fill=white,regular polygon rotate=180]
\tikzstyle{point nosep}=[regular polygon,regular polygon sides=3,draw,scale=0.75,inner sep=-2pt,minimum width=9mm,fill=white,regular polygon rotate=180]
\tikzstyle{copoint}=[regular polygon,regular polygon sides=3,draw,scale=0.75,inner sep=-0.5pt,minimum width=9mm,fill=white]
\tikzstyle{dpoint}=[point,doubled]
\tikzstyle{dcopoint}=[copoint,doubled]

\tikzstyle{pointgrow}=[shape=cornerpoint,kpoint common,scale=0.75,inner sep=3pt]
\tikzstyle{pointgrow dag}=[shape=cornercopoint,kpoint common,scale=0.75,inner sep=3pt]

\tikzstyle{wide copoint}=[fill=white,draw,shape=isosceles triangle,shape border rotate=90,isosceles triangle stretches=true,inner sep=0pt,minimum width=1.5cm,minimum height=6.12mm]
\tikzstyle{wide point}=[fill=white,draw,shape=isosceles triangle,shape border rotate=-90,isosceles triangle stretches=true,inner sep=0pt,minimum width=1.5cm,minimum height=6.12mm,yshift=-0.0mm]
\tikzstyle{wide point plus}=[fill=white,draw,shape=isosceles triangle,shape border rotate=-90,isosceles triangle stretches=true,inner sep=0pt,minimum width=1.74cm,minimum height=7mm,yshift=-0.0mm]

\tikzstyle{wide dpoint}=[fill=white,doubled,draw,shape=isosceles triangle,shape border rotate=-90,isosceles triangle stretches=true,inner sep=0pt,minimum width=1.5cm,minimum height=6.12mm,yshift=-0.0mm]

\tikzstyle{tinypoint}=[regular polygon,regular polygon sides=3,draw,scale=0.55,inner sep=-0.15pt,minimum width=6mm,fill=white,regular polygon rotate=180]

\tikzstyle{white point}=[point]
\tikzstyle{white dpoint}=[dpoint]
\tikzstyle{green point}=[white point] % for backwards-compatibility
\tikzstyle{white copoint}=[copoint]
\tikzstyle{gray point}=[point,fill=gray!40!white]
\tikzstyle{gray dpoint}=[gray point,doubled]
\tikzstyle{red point}=[gray point] % for backwards-compatibility
\tikzstyle{gray copoint}=[copoint,fill=gray!40!white]
\tikzstyle{gray dcopoint}=[gray copoint,doubled]

\tikzstyle{white point guide}=[regular polygon,regular polygon sides=3,font=\scriptsize,draw,scale=0.65,inner sep=-0.5pt,minimum width=9mm,fill=white,regular polygon rotate=180]

\tikzstyle{black point}=[point,fill=black,font=\color{white}]
\tikzstyle{black copoint}=[copoint,fill=black,font=\color{white}]

\tikzstyle{tiny gray point}=[tinypoint,fill=gray!40!white]

\tikzstyle{diredge}=[->]
\tikzstyle{ddiredge}=[<->]
\tikzstyle{rdiredge}=[<-]
\tikzstyle{thickdiredge}=[->, very thick]
\tikzstyle{pointer edge}=[->,very thick,gray]
\tikzstyle{pointer edge part}=[very thick,gray]
\tikzstyle{dashed edge}=[dashed]
\tikzstyle{thick dashed edge}=[very thick,dashed]
\tikzstyle{thick gray dashed edge}=[thick dashed edge,gray!40]
\tikzstyle{thick map edge}=[very thick,|->]

\makeatletter
\newcommand{\boxshape}[3]{%
\pgfdeclareshape{#1}{
\inheritsavedanchors[from=rectangle] % this is nearly a rectangle
\inheritanchorborder[from=rectangle]
\inheritanchor[from=rectangle]{center}
\inheritanchor[from=rectangle]{north}
\inheritanchor[from=rectangle]{south}
\inheritanchor[from=rectangle]{west}
\inheritanchor[from=rectangle]{east}
\backgroundpath{% this is new
\southwest \pgf@xa=\pgf@x \pgf@ya=\pgf@y
\northeast \pgf@xb=\pgf@x \pgf@yb=\pgf@y

\@tempdima=#2
\@tempdimb=#3

\pgfpathmoveto{\pgfpoint{\pgf@xa - 5pt + \@tempdima}{\pgf@ya}}
\pgfpathlineto{\pgfpoint{\pgf@xa - 5pt - \@tempdima}{\pgf@yb}}
\pgfpathlineto{\pgfpoint{\pgf@xb + 5pt + \@tempdimb}{\pgf@yb}}
\pgfpathlineto{\pgfpoint{\pgf@xb + 5pt - \@tempdimb}{\pgf@ya}}
\pgfpathlineto{\pgfpoint{\pgf@xa - 5pt + \@tempdima}{\pgf@ya}}
\pgfpathclose
}
}}

\boxshape{NEbox}{0pt}{3pt}
\boxshape{SEbox}{0pt}{-3pt}
\boxshape{NWbox}{5pt}{0pt}
\boxshape{SWbox}{-5pt}{0pt}
\boxshape{EBox}{-3pt}{3pt}
\boxshape{WBox}{3pt}{-3pt}
\makeatother

\tikzstyle{cloud}=[shape=cloud,draw,minimum width=1.5cm,minimum height=1.5cm]

\tikzstyle{map}=[draw,shape=NEbox,inner sep=1pt,minimum height=4mm,fill=white]
\tikzstyle{dashedmap}=[draw,dashed,shape=NEbox,inner sep=2pt,minimum height=6mm,fill=white]
\tikzstyle{mapdag}=[draw,shape=SEbox,inner sep=1pt,minimum height=4mm,fill=white]
\tikzstyle{mapadj}=[draw,shape=SEbox,inner sep=2pt,minimum height=6mm,fill=white]
\tikzstyle{maptrans}=[draw,shape=SWbox,inner sep=2pt,minimum height=6mm,fill=white]
\tikzstyle{mapconj}=[draw,shape=NWbox,inner sep=2pt,minimum height=6mm,fill=white]

\tikzstyle{medium map}=[draw,shape=NEbox,inner sep=2pt,minimum height=6mm,fill=white,minimum width=7mm]
\tikzstyle{medium map dag}=[draw,shape=SEbox,inner sep=2pt,minimum height=6mm,fill=white,minimum width=7mm]
\tikzstyle{medium map adj}=[draw,shape=SEbox,inner sep=2pt,minimum height=6mm,fill=white,minimum width=7mm]
\tikzstyle{medium map trans}=[draw,shape=SWbox,inner sep=2pt,minimum height=6mm,fill=white,minimum width=7mm]
\tikzstyle{medium map conj}=[draw,shape=NWbox,inner sep=2pt,minimum height=6mm,fill=white,minimum width=7mm]
\tikzstyle{semilarge map}=[draw,shape=NEbox,inner sep=2pt,minimum height=6mm,fill=white,minimum width=9.5mm]
\tikzstyle{semilarge map trans}=[draw,shape=SWbox,inner sep=2pt,minimum height=6mm,fill=white,minimum width=9.5mm]
\tikzstyle{semilarge map adj}=[draw,shape=SEbox,inner sep=2pt,minimum height=6mm,fill=white,minimum width=9.5mm]
\tikzstyle{semilarge map dag}=[draw,shape=SEbox,inner sep=2pt,minimum height=6mm,fill=white,minimum width=9.5mm]
\tikzstyle{semilarge map conj}=[draw,shape=NWbox,inner sep=2pt,minimum height=6mm,fill=white,minimum width=9.5mm]
\tikzstyle{large map}=[draw,shape=NEbox,inner sep=2pt,minimum height=6mm,fill=white,minimum width=12mm]
\tikzstyle{large map conj}=[draw,shape=NWbox,inner sep=2pt,minimum height=6mm,fill=white,minimum width=12mm]
\tikzstyle{very large map}=[draw,shape=NEbox,inner sep=2pt,minimum height=6mm,fill=white,minimum width=17mm]

\tikzstyle{medium dmap}=[draw,doubled,shape=NEbox,inner sep=2pt,minimum height=6mm,fill=white,minimum width=7mm]
\tikzstyle{medium dmap dag}=[draw,doubled,shape=SEbox,inner sep=2pt,minimum height=6mm,fill=white,minimum width=7mm]
\tikzstyle{medium dmap adj}=[draw,doubled,shape=SEbox,inner sep=2pt,minimum height=6mm,fill=white,minimum width=7mm]
\tikzstyle{medium dmap trans}=[draw,doubled,shape=SWbox,inner sep=2pt,minimum height=6mm,fill=white,minimum width=7mm]
\tikzstyle{medium dmap conj}=[draw,doubled,shape=NWbox,inner sep=2pt,minimum height=6mm,fill=white,minimum width=7mm]
\tikzstyle{semilarge dmap}=[draw,doubled,shape=NEbox,inner sep=2pt,minimum height=6mm,fill=white,minimum width=9.5mm]
\tikzstyle{semilarge dmap trans}=[draw,doubled,shape=SWbox,inner sep=2pt,minimum height=6mm,fill=white,minimum width=9.5mm]
\tikzstyle{semilarge dmap adj}=[draw,doubled,shape=SEbox,inner sep=2pt,minimum height=6mm,fill=white,minimum width=9.5mm]
\tikzstyle{semilarge dmap dag}=[draw,doubled,shape=SEbox,inner sep=2pt,minimum height=6mm,fill=white,minimum width=9.5mm]
\tikzstyle{semilarge dmap conj}=[draw,doubled,shape=NWbox,inner sep=2pt,minimum height=6mm,fill=white,minimum width=9.5mm]
\tikzstyle{large dmap}=[draw,doubled,shape=NEbox,inner sep=2pt,minimum height=6mm,fill=white,minimum width=12mm]
\tikzstyle{large dmap conj}=[draw,doubled,shape=NWbox,inner sep=2pt,minimum height=6mm,fill=white,minimum width=12mm]
\tikzstyle{large dmap trans}=[draw,doubled,shape=SWbox,inner sep=2pt,minimum height=6mm,fill=white,minimum width=12mm]
\tikzstyle{large dmap adj}=[draw,doubled,shape=SEbox,inner sep=2pt,minimum height=6mm,fill=white,minimum width=12mm]
\tikzstyle{large dmap dag}=[draw,doubled,shape=SEbox,inner sep=2pt,minimum height=6mm,fill=white,minimum width=12mm]
\tikzstyle{very large dmap}=[draw,doubled,shape=NEbox,inner sep=2pt,minimum height=6mm,fill=white,minimum width=19.5mm]

\tikzstyle{muxbox}=[draw,shape=rectangle,minimum height=3mm,minimum width=3mm,fill=white]
\tikzstyle{dmuxbox}=[muxbox,doubled]

\tikzstyle{box}=[draw,shape=rectangle,inner sep=2pt,minimum height=6mm,minimum width=6mm,fill=white]
\tikzstyle{dbox}=[draw,doubled,shape=rectangle,inner sep=2pt,minimum height=6mm,minimum width=6mm,fill=white]
\tikzstyle{dmap}=[draw,doubled,shape=NEbox,inner sep=2pt,minimum height=6mm,fill=white]
\tikzstyle{dmapdag}=[draw,doubled,shape=SEbox,inner sep=2pt,minimum height=6mm,fill=white]
\tikzstyle{dmapadj}=[draw,doubled,shape=SEbox,inner sep=2pt,minimum height=6mm,fill=white]
\tikzstyle{dmaptrans}=[draw,doubled,shape=SWbox,inner sep=2pt,minimum height=6mm,fill=white]
\tikzstyle{dmapconj}=[draw,doubled,shape=NWbox,inner sep=2pt,minimum height=6mm,fill=white]

\tikzstyle{ddmap}=[draw,doubled,dashed,shape=NEbox,inner sep=2pt,minimum height=6mm,fill=white]
\tikzstyle{ddmapdag}=[draw,doubled,dashed,shape=SEbox,inner sep=2pt,minimum height=6mm,fill=white]
\tikzstyle{ddmapadj}=[draw,doubled,dashed,shape=SEbox,inner sep=2pt,minimum height=6mm,fill=white]
\tikzstyle{ddmaptrans}=[draw,doubled,dashed,shape=SWbox,inner sep=2pt,minimum height=6mm,fill=white]
\tikzstyle{ddmapconj}=[draw,doubled,dashed,shape=NWbox,inner sep=2pt,minimum height=6mm,fill=white]

\boxshape{sNEbox}{0pt}{3pt}
\boxshape{sSEbox}{0pt}{-3pt}
\boxshape{sNWbox}{3pt}{0pt}
\boxshape{sSWbox}{-3pt}{0pt}
\tikzstyle{smap}=[draw,shape=sNEbox,fill=white]
\tikzstyle{smapdag}=[draw,shape=sSEbox,fill=white]
\tikzstyle{smapadj}=[draw,shape=sSEbox,fill=white]
\tikzstyle{smaptrans}=[draw,shape=sSWbox,fill=white]
\tikzstyle{smapconj}=[draw,shape=sNWbox,fill=white]

\tikzstyle{dsmap}=[draw,dashed,shape=sNEbox,fill=white]
\tikzstyle{dsmapdag}=[draw,dashed,shape=sSEbox,fill=white]
\tikzstyle{dsmaptrans}=[draw,dashed,shape=sSWbox,fill=white]
\tikzstyle{dsmapconj}=[draw,dashed,shape=sNWbox,fill=white]

\boxshape{mNEbox}{0pt}{10pt}
\boxshape{mSEbox}{0pt}{-10pt}
\boxshape{mNWbox}{10pt}{0pt}
\boxshape{mSWbox}{-10pt}{0pt}
\tikzstyle{mmap}=[draw,shape=mNEbox]
\tikzstyle{mmapdag}=[draw,shape=mSEbox]
\tikzstyle{mmaptrans}=[draw,shape=mSWbox]
\tikzstyle{mmapconj}=[draw,shape=mNWbox]

\tikzstyle{mmapgray}=[draw,fill=gray!40!white,shape=mNEbox]
\tikzstyle{smapgray}=[draw,fill=gray!40!white,shape=sNEbox]

\makeatletter

\pgfdeclareshape{cornerpoint}{
\inheritsavedanchors[from=rectangle] % this is nearly a rectangle
\inheritanchorborder[from=rectangle]
\inheritanchor[from=rectangle]{center}
\inheritanchor[from=rectangle]{north}
\inheritanchor[from=rectangle]{south}
\inheritanchor[from=rectangle]{west}
\inheritanchor[from=rectangle]{east}
\backgroundpath{% this is new
\southwest \pgf@xa=\pgf@x \pgf@ya=\pgf@y
\northeast \pgf@xb=\pgf@x \pgf@yb=\pgf@y

\pgfmathsetmacro{\pgf@shorten@left}{\pgfkeysvalueof{/tikz/shorten left}}
\pgfmathsetmacro{\pgf@shorten@right}{\pgfkeysvalueof{/tikz/shorten right}}

\pgfpathmoveto{\pgfpoint{0.5 * (\pgf@xa + \pgf@xb)}{\pgf@ya - 5pt}}
\pgfpathlineto{\pgfpoint{\pgf@xa - 8pt + \pgf@shorten@left}{\pgf@yb - 1.5 * \pgf@shorten@left}}
\pgfpathlineto{\pgfpoint{\pgf@xa - 8pt + \pgf@shorten@left}{\pgf@yb}}
\pgfpathlineto{\pgfpoint{\pgf@xb + 8pt - \pgf@shorten@right}{\pgf@yb}}
\pgfpathlineto{\pgfpoint{\pgf@xb + 8pt - \pgf@shorten@right}{\pgf@yb - 1.5 * \pgf@shorten@right}}
\pgfpathclose
}
}

\pgfdeclareshape{cornercopoint}{
\inheritsavedanchors[from=rectangle] % this is nearly a rectangle
\inheritanchorborder[from=rectangle]
\inheritanchor[from=rectangle]{center}
\inheritanchor[from=rectangle]{north}
\inheritanchor[from=rectangle]{south}
\inheritanchor[from=rectangle]{west}
\inheritanchor[from=rectangle]{east}
\backgroundpath{% this is new
\southwest \pgf@xa=\pgf@x \pgf@ya=\pgf@y
\northeast \pgf@xb=\pgf@x \pgf@yb=\pgf@y

\pgfmathsetmacro{\pgf@shorten@left}{\pgfkeysvalueof{/tikz/shorten left}}
\pgfmathsetmacro{\pgf@shorten@right}{\pgfkeysvalueof{/tikz/shorten right}}

\pgfpathmoveto{\pgfpoint{0.5 * (\pgf@xa + \pgf@xb)}{\pgf@yb + 5pt}}
\pgfpathlineto{\pgfpoint{\pgf@xa - 8pt + \pgf@shorten@left}{\pgf@ya + 1.5 * \pgf@shorten@left}}
\pgfpathlineto{\pgfpoint{\pgf@xa - 8pt + \pgf@shorten@left}{\pgf@ya}}
\pgfpathlineto{\pgfpoint{\pgf@xb + 8pt - \pgf@shorten@right}{\pgf@ya}}
\pgfpathlineto{\pgfpoint{\pgf@xb + 8pt - \pgf@shorten@right}{\pgf@ya + 1.5 * \pgf@shorten@right}}
\pgfpathclose
}
}

\makeatother

\pgfkeyssetvalue{/tikz/shorten left}{0pt}
\pgfkeyssetvalue{/tikz/shorten right}{0pt}

\tikzstyle{kpoint common}=[draw,fill=white,inner sep=1pt,minimum height=4mm]
\tikzstyle{kpoint sc}=[shape=cornerpoint,kpoint common]
\tikzstyle{kpoint adjoint sc}=[shape=cornercopoint,kpoint common]
\tikzstyle{kpoint}=[shape=cornerpoint,shorten left=5pt,kpoint common]
\tikzstyle{kpoint adjoint}=[shape=cornercopoint,shorten left=5pt,kpoint common]
\tikzstyle{kpoint conjugate}=[shape=cornerpoint,shorten right=5pt,kpoint common]
\tikzstyle{kpoint transpose}=[shape=cornercopoint,shorten right=5pt,kpoint common]
\tikzstyle{kpoint symm}=[shape=cornerpoint,shorten left=5pt,shorten right=5pt,kpoint common]

\tikzstyle{wide kpoint sc}=[shape=cornerpoint,kpoint common, minimum width=1 cm]
\tikzstyle{wide kpointdag sc}=[shape=cornercopoint,kpoint common, minimum width=1 cm]

\tikzstyle{black kpoint}=[shape=cornerpoint,shorten left=5pt,kpoint common,fill=black,font=\color{white}]

\tikzstyle{black kpoint sm}=[shape=cornerpoint,shorten left=5pt,kpoint common,fill=black,font=\color{white},scale=0.75]

\tikzstyle{black kpoint adjoint}=[shape=cornercopoint,shorten left=5pt,kpoint common,fill=black,font=\color{white}]
\tikzstyle{black kpointadj}=[shape=cornercopoint,shorten left=5pt,kpoint common,fill=black,font=\color{white}]

\tikzstyle{black kpointadj sm}=[shape=cornercopoint,shorten left=5pt,kpoint common,fill=black,font=\color{white},scale=0.75]

\tikzstyle{black dkpoint}=[shape=cornerpoint,shorten left=5pt,kpoint common,fill=black, doubled,font=\color{white}]
\tikzstyle{black dkpoint adjoint}=[shape=cornercopoint,shorten left=5pt,kpoint common,fill=black, doubled,font=\color{white}]
\tikzstyle{black dkpointadj}=[shape=cornercopoint,shorten left=5pt,kpoint common,fill=black, doubled,font=\color{white}]

\tikzstyle{black dkpoint sm}=[shape=cornerpoint,shorten left=5pt,kpoint common,fill=black, doubled,font=\color{white},scale=0.75]
\tikzstyle{black dkpointadj sm}=[shape=cornercopoint,shorten left=5pt,kpoint common,fill=black, doubled,font=\color{white},scale=0.75]

\tikzstyle{kpointdag}=[kpoint adjoint]
\tikzstyle{kpointadj}=[kpoint adjoint]
\tikzstyle{kpointconj}=[kpoint conjugate]
\tikzstyle{kpointtrans}=[kpoint transpose]

\tikzstyle{big kpoint}=[kpoint, minimum width=1.2 cm, minimum height=8mm, inner sep=4pt, text depth=3mm]

\tikzstyle{wide kpoint}=[kpoint, minimum width=1 cm, inner sep=2pt]%, text depth=-0.7 mm]
\tikzstyle{wide kpointdag}=[kpointdag, minimum width=1 cm, inner sep=2pt]%, text depth=0.7 mm]
\tikzstyle{wide kpointconj}=[kpointconj, minimum width=1 cm, inner sep=2pt]%, text depth=-0.7 mm]
\tikzstyle{wide kpointtrans}=[kpointtrans, minimum width=1 cm, inner sep=2pt]%, text depth=0.7 mm]

\tikzstyle{wider kpoint}=[kpoint, minimum width=1.25 cm, inner sep=2pt]%, text depth=-0.7 mm]
\tikzstyle{wider kpointdag}=[kpointdag, minimum width=1.25 cm, inner sep=2pt]%, text depth=0.7 mm]
\tikzstyle{wider kpointconj}=[kpointconj, minimum width=1.25 cm, inner sep=2pt]%, text depth=-0.7 mm]
\tikzstyle{wider kpointtrans}=[kpointtrans, minimum width=1.25 cm, inner sep=2pt]%, text depth=0.7 mm]

\tikzstyle{gray kpoint}=[kpoint,fill=gray!50!white]
\tikzstyle{gray kpointdag}=[kpointdag,fill=gray!50!white]
\tikzstyle{gray kpointadj}=[kpointadj,fill=gray!50!white]
\tikzstyle{gray kpointconj}=[kpointconj,fill=gray!50!white]
\tikzstyle{gray kpointtrans}=[kpointtrans,fill=gray!50!white]

\tikzstyle{gray dkpoint}=[kpoint,fill=gray!50!white,doubled]
\tikzstyle{gray dkpointdag}=[kpointdag,fill=gray!50!white,doubled]
\tikzstyle{gray dkpointadj}=[kpointadj,fill=gray!50!white,doubled]
\tikzstyle{gray dkpointconj}=[kpointconj,fill=gray!50!white,doubled]
\tikzstyle{gray dkpointtrans}=[kpointtrans,fill=gray!50!white,doubled]

\tikzstyle{white label}=[draw,fill=white,rectangle,inner sep=0.7 mm]
\tikzstyle{gray label}=[draw,fill=gray!50!white,rectangle,inner sep=0.7 mm]
\tikzstyle{black label}=[draw,fill=black,rectangle,inner sep=0.7 mm]

\tikzstyle{dkpoint}=[kpoint,doubled]
\tikzstyle{wide dkpoint}=[wide kpoint,doubled]
\tikzstyle{dkpointdag}=[kpoint adjoint,doubled]
\tikzstyle{wide dkpointdag}=[wide kpointdag,doubled]
\tikzstyle{dkcopoint}=[kpoint adjoint,doubled]
\tikzstyle{dkpointadj}=[kpoint adjoint,doubled]
\tikzstyle{dkpointconj}=[kpoint conjugate,doubled]
\tikzstyle{dkpointtrans}=[kpoint transpose,doubled]

\tikzstyle{kscalar}=[kpoint common, shape=EBox, inner xsep=-1pt, inner ysep=3pt,font=\small]
\tikzstyle{kscalarconj}=[kpoint common, shape=WBox, inner xsep=-1pt, inner ysep=3pt,font=\small]

\tikzstyle{spekpoint}=[kpoint sc,minimum height=5mm,inner sep=3pt]
\tikzstyle{spekcopoint}=[kpoint adjoint sc,minimum height=5mm,inner sep=3pt]

\tikzstyle{dspekpoint}=[spekpoint,doubled]
\tikzstyle{dspekcopoint}=[spekcopoint,doubled]

 \tikzstyle{upground}=[circuit ee IEC,thick,ground,rotate=90,scale=2.5]
 \tikzstyle{downground}=[circuit ee IEC,thick,ground,rotate=-90,scale=2.5]
 \tikzstyle{bigground}=[circuit ee IEC,thick,ground,rotate=90,xscale=2.5,yscale=4.5]
\tikzstyle{arrs}=[-latex,font=\small,auto]
\tikzstyle{arrow plain}=[arrs]
\tikzstyle{arrow dashed}=[dashed,arrs]
\tikzstyle{arrow bold}=[very thick,arrs]
\tikzstyle{arrow hide}=[draw=white!0,-]
\tikzstyle{arrow reverse}=[latex-]
\tikzstyle{cdnode}=[]

\let\olddagger\dagger
\renewcommand{\dagger}{\ensuremath{\olddagger}\xspace}

\theoremstyle{definition}
\newtheorem{theorem}{Theorem}[]

\newtheorem{definition}[]{Definition}

\newtheorem{example*}[theorem]{Example*}
\newtheorem{examples*}[theorem]{Examples*}

\newtheorem{remark*}[theorem]{Remark*}

\theoremstyle{plain}

\usepackage{color}
\def\bR{\begin{color}{red}}
\def\bB{\begin{color}{blue}}
\def\bM{\begin{color}{magenta}}
\def\bC{\begin{color}{cyan}}
\def\bW{\begin{color}{white}}
\def\bBl{\begin{color}{black}}
\def\bG{\begin{color}{green}}
\def\bY{\begin{color}{yellow}}
\def\e{\end{color}\xspace}
\newcommand{\bit}{\begin{itemize}}
\newcommand{\eit}{\end{itemize}\par\noindent}
\newcommand{\ben}{\begin{enumerate}}
\newcommand{\een}{\end{enumerate}\par\noindent}
\newcommand{\beq}{\begin{equation}}
\newcommand{\eeq}{\end{equation}\par\noindent}
\newcommand{\beqa}{\begin{eqnarray*}}
\newcommand{\eeqa}{\end{eqnarray*}\par\noindent}
\newcommand{\beqn}{\begin{eqnarray}}
\newcommand{\eeqn}{\end{eqnarray}\par\noindent}
\def\jR{\begin{color}{DarkRed}}
\def\jB{\begin{color}{MidnightBlue}}
\def\jM{\begin{color}{magenta}}
\def\jC{\begin{color}{cyan}}
\def\jW{\begin{color}{white}}
\def\jBl{\begin{color}{black}}
\def\jG{\begin{color}{green}}
\def\jY{\begin{color}{yellow}}

\def\cR{\begin{color}{Crimson}}
\def\cB{\begin{color}{cyan}}
\def\cM{\begin{color}{magenta}}
\def\cC{\begin{color}{cyan}}
\def\cW{\begin{color}{white}}
\def\cBl{\begin{color}{black}}
\def\cG{\begin{color}{green}}
\def\cY{\begin{color}{yellow}}
\usepackage{enumitem}
\setlist[itemize]{noitemsep}
\setlist[enumerate]{noitemsep}

\begin{document}

\title{A no-go theorem on the nature of the gravitational field beyond quantum theory}
\date{03-08-2022}
\author{Thomas D. Galley${}^\mathsection$}
\email{tgalley1@perimeterinstitute.ca}
\affiliation{Perimeter Institute for Theoretical Physics, 31 Caroline St. N, Waterloo, Ontario, N2L 2Y5, Canada
}

\author{Flaminia Giacomini${}^\mathsection$}
 \email{fgiacomini@perimeterinstitute.ca}
\affiliation{Perimeter Institute for Theoretical Physics, 31 Caroline St. N, Waterloo, Ontario, N2L 2Y5, Canada
}

\author{John H. Selby${}^\mathsection$}
 \email{john.h.selby@gmail.com}
\affiliation{ICTQT, University of Gda\'{n}sk, Wita Stwosza 63, 80-308 Gda\'{n}sk, Poland
}

\begin{abstract}
Recently, table-top experiments involving massive quantum systems have been proposed to test the interface of quantum theory and gravity. In particular, the crucial point of the debate is whether it is possible to conclude anything on the quantum nature of the gravitational field, provided that two quantum systems become entangled solely due to the gravitational interaction. Typically, this question has been addressed by assuming a specific physical theory to describe the gravitational interaction, but no systematic approach to characterise the set of possible gravitational theories which are compatible with the observation of entanglement has been proposed. Here, we remedy this by introducing the framework of Generalised Probabilistic Theories (GPTs) to the study of the nature of the gravitational field. This framework enables us to systematically study all theories compatible with the detection of entanglement generated via the gravitational interaction between two systems. We prove a no-go theorem stating that the following statements are incompatible: i) gravity is able to generate entanglement; ii) gravity mediates the interaction between the systems; iii) gravity is classical. We analyse the violation of each condition, in particular with respect to alternative non-linear models such as the Schr{\"o}dinger-Newton equation and Collapse Models.

\end{abstract}

\maketitle

\def\thefootnote{$\mathsection$}\footnotetext{All authors contributed equally to this work.}

\section{Introduction}

The fundamental description of spacetime and the nature of the gravitational field are among the deepest puzzles that fundamental physics is currently facing. The view supported by the majority of results is in favour of some sort of quantisation of the gravitational field, however, a generally accepted formulation of a quantum theory of gravity, and the connection to possible experiments, is still lacking. A new opportunity could be opened soon thanks to recent developments in quantum technologies, in particular thanks to the possibility of measuring the properties of the gravitational field associated to a quantum source in a spatial superposition. In light of this effort, it is a pressing question now to determine precisely what such low-energy experiments can teach us about the nature of the gravitational field. The relevance of such low-energy tests to the formulation of physics at the interface between quantum theory and gravity was first discussed by Feynman in 1957~\cite{cecile2011role, zeh2011feynman} and has, since then, attracted a lot of attention~\cite{ford1982gravitational, lindner2005testing, kafri2013noise, kafri2014classical, altamirano2018gravity, anastopoulos2015probing, anastopoulos2020quantum, carlesso2017cavendish, bahrami2015gravity, belenchia2018quantum, belenchia2019information, christodoulou2019possibility, howl2020testing, marshman2020locality, krisnanda2020observable}. Recently, the debate has been revived by two papers~\cite{bose2017spin, marletto2017gravitationally}, in which two masses, $\mathsf{A}$ and $\mathsf{B}$, become entangled via the Newtonian potential. The critical point of the debate \cite{hall2018two, anastopoulos2018comment} is whether any conclusions can be drawn on the quantum nature of the gravitational field, provided that entanglement between the masses is indeed measured as predicted. There is no agreement on this point: some authors~\cite{bose2017spin, marletto2017gravitationally, christodoulou2019possibility} consider this as a genuine quantum feature of gravity, others~\cite{hall2018two, anastopoulos2018comment} argue that this experiment is not conclusive. However, the vast majority of the literature (an exception are Refs.~\cite{marletto2017we, marletto2020witnessing}) assumes an underlying (quantum) model of the gravitational field, so at best these results can show that the experiment is compatible with a quantum treatment of gravity, but cannot answer the question of whether or not gravity is quantum. 

Here, we approach this question from a theory-independent perspective. Specifically, we derive a no-go theorem on the nature of the gravitational field without assuming any specific description of gravity. Our result is based on operational considerations derived from measurements performed on the matter systems interacting gravitationally (and not directly on the gravitational field). To achieve this, we introduce the tools of Generalised Probabilistic Theories (GPTs) to the study of the gravitational field. GPTs have been useful for a large number of problems in thermodynamics \cite{barnum2010entropy,
chiribella2015entanglement,
chiribella2017microcanonical,
barnum2015entropy},
interference \cite{
lee2017higher,
garner2018interferometric,
barnum2014higher,
dakic2014density,
lee2017higher,
barnum2017ruling,
horvat2020interference},
decoherence \cite{richens2017entanglement,
lee2018no,
scandolo2018possible,selby2017leaks}, 
computation \cite{
lee2015computation,
barrett2017computational,
lee2015computation,
krumm2018quantum,
lee2016generalised,
lee2016deriving,
barnum2018oracles,
muller2012structure},
cryptography \cite{
sikora2018simple,
selby2018make,
sikora2019impossibility,
barnum2008nonclassicality,
lami2018ultimate,
barrett2005no},
contextuality \cite{
schmid2020structure,
schmid2020characterization,
shahandeh2019contextuality,
chiribella2014measurement},
information processing \cite{
bae2016structure,
barnum2011information,
barrett2007information,
JP17,
barnum2007generalized,
barnum2011information,
barnum2012teleportation,
barnum2013ensemble,
heinosaari2019no}, 
correlations \cite{
czekaj2018bell,
barnum2010local,
czekaj2020correlations,
henson2014theory,
weilenmann2020analysing,
lami2018non,
cavalcanti2021witworld} and the black hole information problem~\cite{Muller_2012}
as well as to deepen our understanding of the general structure of physical theories \cite{
masanes2019measurement,
galley2017classification,
galley2018any,
Galley2021howdynamics,
chiribella2011informational,
chiribella2014dilation,
chiribella2014distinguishability,
barnum2014local,
wilce2009four,
wilce2018royal,
barnum2016composites,
barnum2013symmetry,
wilce2011symmetry,
masanes2011derivation,
masanes2013existence,
mueller2013three}. GPTs do not presuppose the quantum formalism but instead study these phenomena in a theory-agnostic way, making operational, potentially probabilistic, predictions about the outcomes of experiments. This does not mean that a theory cannot, and should not, do more than this by providing an `ontological' level of description, but that the operational level of description is a necessary prerequisite of any good physical theory. It is this operational level of description which is captured by the framework of GPTs. One imposes simple mathematical axioms on the structure of the theory, which faithfully capture the conceptual operational underpinnings, and, by doing so, one can derive a rich mathematical structure which any GPT must have. For example, this means that we are no longer free to propose arbitrary modifications of quantum theory, but, having proposed a particular modification, we must then check that it is indeed compatible with the basic mathematical axioms -- for example, that probabilistic predictions must be valued in the interval $[0,1]$. 

In this paper, we apply the tools of GPTs to prove a no-go theorem on the nature of the gravitational field. Interestingly, our no-go theorem allows us to overcome the dichotomy that gravity is either classical or quantum. In particular, this means that we can apply our result to constrain a much richer set of theories, which are neither classical nor quantum, such as hybrid models and alternative modifications of quantum theory, for instance the Schr{\"o}dinger-Newton equation and Collapse Models. \\

\section{Generalised probabilistic theories}
\label{sec:GenPhysT}

Classical and quantum theories are instances of GPTs. We will give a brief sketch of this formalism here, but direct the reader to modern reviews such as Refs.~\cite{chiribella2016quantum,lami2018non,plavala2021general,muller2021probabilistic} for more details. A GPT is a theory of systems with states given by elements of a convex subset of a real vector space, effects correspond to elements of a convex subset of the dual space and transformations induce linear maps which take the set of states to itself. The composition rule is given by a bilinear map from two systems to a third, and the probability of an outcome given a state and an effect is computed via evaluation of the effect on the state. 

For example, in quantum theory the real vector space associated to a system is the space of Hermitian operators over $\C^d$. States are described by density operators, effects by positive semi-definite operators and transformations by completely positive trace preserving channels. Quantum systems compose via the tensor product, and probabilities of measurement outcomes are computed via the standard trace rule. 

The key to our work, however, is the definition of classicality within the GPT framework:

\begin{definition}[Classicality] \label{def:classical}
A system is classical if
\begin{enumerate}
	\item the state space is a simplex, i.e., the space of probability distributions $\lbrace p(x), x\in X \rbrace$, with $X$ being the sample space;
	\item the effect space is a hypercube, given by the functions $X \to [0,1]$. Together with 1. these form the \emph{kinematics} of the classical system.
	\item it composes with the standard tensor product rule with other systems (\emph{composition}).
\end{enumerate}
\end{definition}

Observe that two classical systems $\A$ and $\B$ with sample spaces $X_\A$ and $X_\B$ compose to a classical system with sample space $X_\A \times X_\B$, and so compose via the Cartesian product at the level of sample spaces. However the state spaces (simplices) are embedded in the real vector spaces $\R^{|X_\A|}$, $\R^{|X_\B|}$ and $\R^{|X_\A \times X_\B|}$ respectively. Since $\R^{|X_\A \times X_\B|}=\R^{|X_\A||X_\B|} \simeq \R^{|X_\A|} \otimes \R^{|X_\B|}$  classical systems compose via the tensor product of the underlying real vector spaces.

The key property of classical systems that we will rely on for the proof is that they admit of a resolution of the identity transformation. Let us denote the delta function distributions for $X$ as $\delta_x$ and the atomic effects $\epsilon_x$, which are defined such that $\epsilon_{x'}(\delta_x) = \delta_{x,x'}$, then we have a resolution of the identity transformation given by $\mathds{1}_X = \sum_{x\in X} \delta_x\circ \epsilon_x$. 
This resolution of the identity is a consequence of the fact that classical systems can (at least in principle) be characterised by non-disturbing measurements. The effects associated to these measurements are the atomic effects $\epsilon_x$ in the resolution of identity.

Examples of GPTs beyond classical and quantum theory are discussed in the following. Any GPT which does not obey the definition of classicality given above is said to be a non-classical GPT. In particular, quantum theory is an example of a non-classical theory.

\section{A no-go theorem on the nature of the gravitational field}
\label{sec:NoGoTh}
\begin{figure}[h]
	\begin{center}	
		\includegraphics[scale=0.3]{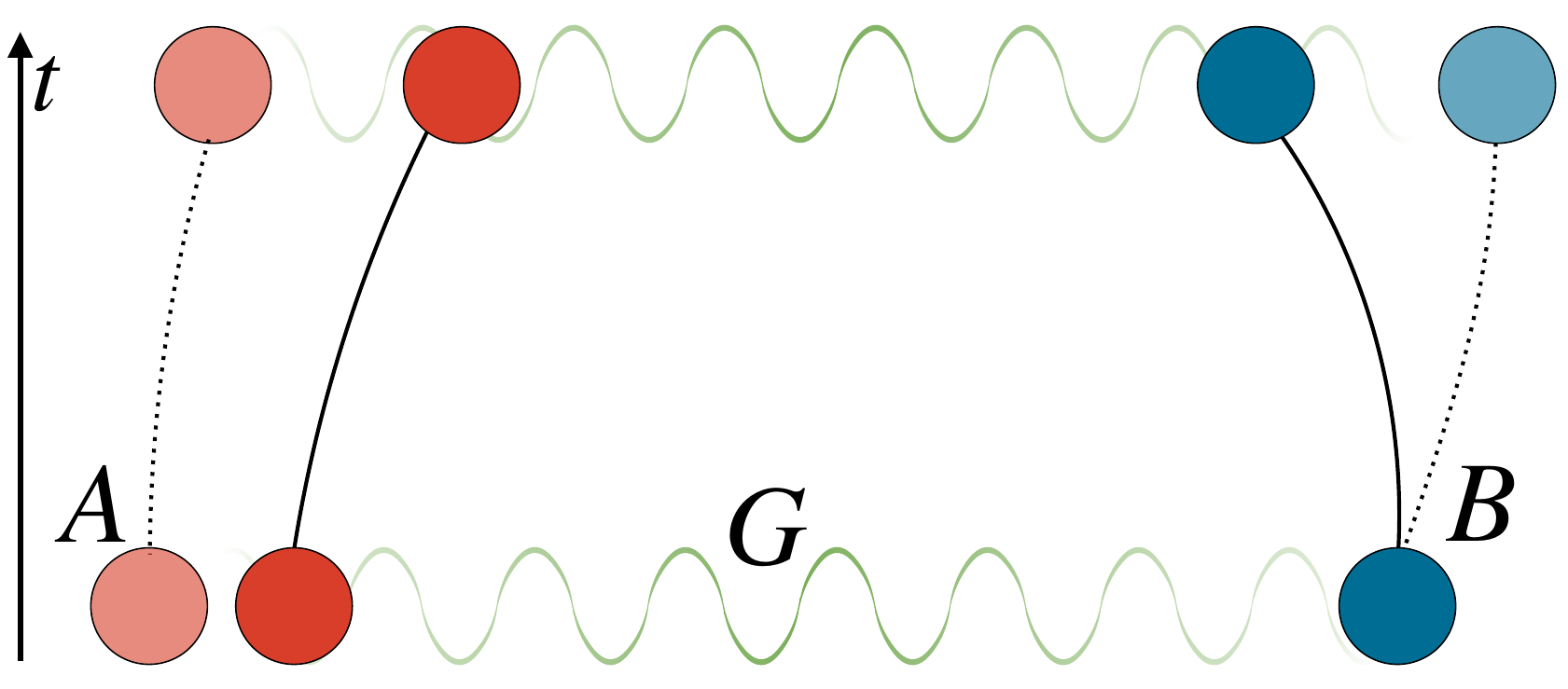}	
		\caption{\label{fig:setup}Illustration of the experimental situation. Two masses $\A$ and $\B$ are initially prepared in a separable state (where at least one of them, $\A$ in the figure, is in a superposition state in position basis). The masses interact via the gravitational field $\G$. After some time, the full state becomes entangled.}	
	\end{center}	
\end{figure}

Consider the following experimental set up, illustrated in Fig.~\ref{fig:setup}.

 We have two distant physical systems $\A$ and $\B$ (the two masses), each of which interacts with some system $\G$ (gravity). We assume that it is possible to experimentally verify that the only way in which they interact is gravitationally, for example, by making sure that all other possible interactions are suitably screened off. For such an experimental set up we prove the following theorem:

\begin{theorem}\label{thm:no-go}
	We consider two GPT systems $\A$ and $\B$, initially in a separable state, and the gravitational field $\G$ in a product state. We assume that the systems $\A$ and $\B$ only interact gravitationally. Then the following statements are incompatible:
	\begin{enumerate}
		\item The gravitational field $\G$ is able to generate entanglement;
		\item $\A$ and $\B$ interact via the mediator $\G$;
		\item $\G$ is classical.
	\end{enumerate}
\end{theorem}
\proof
Here we sketch the proof, for more details see Appendix~\ref{proof:ClassIntTwoSys}. In order to prove the Theorem, we use a generalisation to GPTs of the argument that Local Operations and Classical Communication (LOCC) does not generate entanglement. 

We write the most general interaction which is mediated by a GPT system $\G$ as a chain (possibly infinite) of different rounds of interactions between $\A$ and $\G$, followed by an interaction between $\B$ and $\G$. We denote the interaction between $\A$ and $\G$ at the $n$-th round as $\mathcal{I}_A^{(n)}$, and the interaction between $\B$ and $\G$ at the $n$-th round as $\mathcal{I}_B^{(n)}$. We can then focus on a single one of such rounds. For each round, we insert a resolution of the identity for $\G$ (which is possible if and only if $\G$ is classical) between $\mathcal{I}_A^{(n)}$ and $\mathcal{I}_B^{(n)}$ and post-select on a final state of $\G$. This procedure is equivalent to applying two completely positive trace non-increasing maps to the systems $\A$ and $\B$, which are known to not generate entanglement.
\endproof

We explicitly proved that if conditions 2 and 3 hold, condition 1 is violated.
Hence at least one of the conditions must be violated. We now discuss each of these conditions in turn.

\textsc{Condition 1.} Entanglement generation via gravitational interaction is predicted by most theories applying the superposition principles to the low-energy regime of gravity. However, this fact has not been tested experimentally, and there exist alternative theories~\cite{diosi1984gravitation, ghirardi1986unified, diosi1987universal, ghirardi1990Markov, penrose1996gravity, bassi2003dynamical, Adler_2007, blencowe2013effective, Anastopoulos:2013zya, penrose2014gravitization, bassi2017gravitational} predicting a decoherence or reduction mechanism when massive bodies are put into a spatial superposition. In this case, no generation of entanglement could arise. We explicitly discuss these models in the context of our no-go theorem in the following.

An important subtlety on the meaning of ``entanglement generation'' in this context was first noted in Ref.~\cite{pal2021experimental}, where, in particular, care must be taken in the case that the systems are not initially in a product state. We   make the assumption that our particles are initially in a product state, however, this could easily be generalised following the analysis of Ref.~\cite{pal2021experimental} provided that a suitable measure of entanglement generation is adopted.

\textsc{Condition 2.} In condition 2, by \emph{mediator} we mean that $\A$ and $\B$ do not interact directly, but only via $\G$. In a circuit diagram notation this means that the circuit decomposes into a sequence of transformations on $\A\G$ and $\B\G$. 
Consider a field (such as the electromagnetic field) acting on two physical systems; there are multiple ways of modelling this as a GPT circuit/causal structure: 
	\begin{enumerate}
		\item The field acts non-locally on both systems, for instance as a tranformation on $\A\B$ conditioned on $\G$. In this case condition 2 is not met:
		\beq %
\begin{tikzpicture}
	\begin{pgfonlayer}{nodelayer}
		\node [style=none] (0) at (-1.25, -1.25) {};
		\node [style=none] (1) at (-1.25, -0.5) {};
		\node [style=right label] (2) at (-1.25, -1) {$\A$};
		\node [style=none] (3) at (1.75, -1.25) {};
		\node [style=none] (4) at (1.75, 1.25) {};
		\node [style=right label] (5) at (0.25, 1) {$\B$};
		\node [style=none] (6) at (0.25, -0.5) {};
		\node [style=none] (7) at (0.25, -1.25) {};
		\node [style=none] (9) at (-1.75, -0.5) {};
		\node [style=none] (10) at (-1.75, 0.5) {};
		\node [style=none] (11) at (0.75, 0.5) {};
		\node [style=none] (12) at (0.75, -0.5) {};
		\node [style=none] (13) at (-1.25, 0.5) {};
		\node [style=none] (14) at (-1.25, 1.25) {};
		\node [style=right label] (15) at (-1.25, 1) {$\A$};
		\node [style=none] (31) at (-1.25, 1.25) {};
		\node [style=none] (81) at (0.25, 1.25) {};
		\node [style=none] (82) at (0.25, 0.5) {};
		\node [style=right label] (83) at (1.75, -1) {$\G$};
		\node [style=right label] (84) at (0.25, -1) {$\B$};
		\node [style=black dot] (85) at (1.75, 0) {};
		\node [style=none] (86) at (0.75, 0) {};
		\node [style=right label] (87) at (1.75, 1) {$\G$};
	\end{pgfonlayer}
	\begin{pgfonlayer}{edgelayer}
		\draw [qWire] (1.center) to (0.center);
		\draw [cWire] (4.center) to (3.center);
		\draw [qWire] (6.center) to (7.center);
		\draw (10.center) to (11.center);
		\draw (11.center) to (12.center);
		\draw (12.center) to (9.center);
		\draw (9.center) to (10.center);
		\draw [qWire] (14.center) to (13.center);
		\draw [qWire] (81.center) to (82.center);
		\draw (86.center) to (85);
	\end{pgfonlayer}
\end{tikzpicture}}.\eeq
		\item The field $\G$ acts locally on $\A$ and $\B$ and can be modeled as a product of three GPT systems $\G_\A$, $\G_\B$ and $\G_R$, with $\G_A$ being the part of the field which is local to $\A$; $\G_\B$ being the part which is local to $\B$ and $\G_R$ being the part of the field which does not interact directly with $\A$ or $\B$. Here an interaction between $\A$ and $\B$ which is mediated by $\G$ would consist of a sequence of interactions between $\A \G_\A$, $\B \G_\B$, $\G_\A \G_\B$ and $\G_\A \G_R \G_\B$:\beq
\begin{tikzpicture}
	\begin{pgfonlayer}{nodelayer}
		\node [style=none] (0) at (-1.75, -2) {};
		\node [style=none] (1) at (-1.75, -1.25) {};
		\node [style=right label] (2) at (-1.75, -1.75) {$\A$};
		\node [style=none] (3) at (3.25, -2) {};
		\node [style=none] (4) at (3.25, -1.25) {};
		\node [style=right label] (5) at (3.25, -1.75) {$\B$};
		\node [style=none] (6) at (-0.25, -1.25) {};
		\node [style=none] (7) at (-0.25, -2) {};
		\node [style=right label] (8) at (-0.25, -1.75) {$\G_A$};
		\node [style=none] (9) at (-2.25, -1.25) {};
		\node [style=none] (10) at (-2.25, -0.25) {};
		\node [style=none] (11) at (0.25, -0.25) {};
		\node [style=none] (12) at (0.25, -1.25) {};
		\node [style=none] (13) at (-1.75, -0.25) {};
		\node [style=none] (14) at (-1.75, 2.25) {};
		\node [style=right label] (15) at (-1.75, 0) {$\A$};
		\node [style=none] (19) at (1.25, -1.25) {};
		\node [style=none] (20) at (1.25, -0.25) {};
		\node [style=none] (21) at (3.75, -0.25) {};
		\node [style=none] (22) at (3.75, -1.25) {};
		\node [style=none] (25) at (3.25, 2.25) {};
		\node [style=none] (26) at (3.25, -0.25) {};
		\node [style=right label] (28) at (3.25, 0) {$\B$};
		\node [style=none] (31) at (-1.75, 2.25) {};
		\node [style=none] (75) at (0.75, 0.5) {};
		\node [style=none] (76) at (0.75, -2) {};
		\node [style=right label] (77) at (0.75, -1.75) {$\G_R$};
		\node [style=none] (78) at (1.75, -1.25) {};
		\node [style=none] (79) at (1.75, -2) {};
		\node [style=right label] (80) at (1.75, -1.75) {$\G_B$};
		\node [style=none] (81) at (-0.25, 0.5) {};
		\node [style=none] (82) at (-0.25, -0.25) {};
		\node [style=right label] (83) at (-0.25, 0) {$\G_A$};
		\node [style=none] (84) at (1.75, 0.5) {};
		\node [style=none] (85) at (1.75, -0.25) {};
		\node [style=right label] (86) at (1.75, 0) {$\G_B$};
		\node [style=none] (88) at (-0.75, 0.5) {};
		\node [style=none] (89) at (-0.75, 1.5) {};
		\node [style=none] (90) at (2.25, 1.5) {};
		\node [style=none] (91) at (2.25, 0.5) {};
		\node [style=none] (92) at (-0.25, 2.25) {};
		\node [style=none] (93) at (-0.25, 1.5) {};
		\node [style=right label] (94) at (-0.25, 1.75) {$\G_A$};
		\node [style=none] (100) at (0.75, 1.5) {};
		\node [style=right label] (101) at (0.75, 1.75) {$\G_R$};
		\node [style=none] (102) at (1.75, 2.25) {};
		\node [style=none] (103) at (1.75, 1.5) {};
		\node [style=right label] (104) at (1.75, 1.75) {$\G_B$};
		\node [style=none] (105) at (0.75, 2.25) {};
		\node [style=none] (106) at (0.75, 1.5) {};
	\end{pgfonlayer}
	\begin{pgfonlayer}{edgelayer}
		\draw [qWire] (1.center) to (0.center);
		\draw [qWire] (4.center) to (3.center);
		\draw (6.center) to (7.center);
		\draw (10.center) to (11.center);
		\draw (11.center) to (12.center);
		\draw (12.center) to (9.center);
		\draw (9.center) to (10.center);
		\draw [qWire] (14.center) to (13.center);
		\draw (20.center) to (21.center);
		\draw (21.center) to (22.center);
		\draw (22.center) to (19.center);
		\draw (19.center) to (20.center);
		\draw [qWire] (25.center) to (26.center);
		\draw (75.center) to (76.center);
		\draw (78.center) to (79.center);
		\draw (81.center) to (82.center);
		\draw (84.center) to (85.center);
		\draw (89.center) to (90.center);
		\draw (90.center) to (91.center);
		\draw (91.center) to (88.center);
		\draw (88.center) to (89.center);
		\draw (92.center) to (93.center);
		\draw (102.center) to (103.center);
		\draw (105.center) to (106.center);
	\end{pgfonlayer}
\end{tikzpicture}}.
\eeq
	\end{enumerate}

Observe that the theorem does not require that $\A$ and $\B$ are space-like separated, since it requires only that $\G$ mediates the interaction. However, a concrete experimental test of the theorem would require the local operations on $\A$ and $\B$ to be performed at space-like distances to enforce and verify Condition 2. Notice that the violation of Condition 2 allows the interaction to be a direct interaction between $\A$ and $\B$. In the case 1. where the field is not a mediator, the treatment of the gravitational field is operationally indistinguishable from the direct interaction between $\A$ and $\B$, unless one performs measurements directly on $\G$ (which we assume is not possible). In quantum theory direct interactions between $\A$ and $\B$ can lead to faster than light signalling when the two systems $\A\B$ are space-like separated, as shown in Appendix~\ref{sec:GravityMed}. Therefore by ensuring that $\A$ and $\B$ are space-like separated the assumption of no-faster than light signalling guarantees that Condition 2. is met.

\textsc{Condition 3.} When condition 3 is violated, $\G$ is a non-classical system. Crucially observe that this does not imply that $\G$ is quantum, and allows for a ``hybrid'' description of $\A$, $\B$, and $\G$, where for instance $\A$ and $\B$ are quantum systems, while $\G$ is neither classical nor quantum. The generality of our result is particularly interesting in light of two considerations: i) the fact that $\A$ and $\B$ are not assumed to be quantum ensures the robustness of our result against modifications of quantum theory, i.e., in a more general theory, which recovers quantum theory as a limit; ii) in the case where $\A$ and $\B$ are quantum, the use of the GPT framework opens the possibility to find some physical assumptions that uniquely constrain the gravitational field to be quantum from the scope of all non-classical possibilities.

Concerning (i), we identify two cases in which $\A$ and $\B$ might not be quantum systems. The first one can occur if there exist measurements which are not technically achievable currently, but hypothetically may be with access to
more highly sensitive devices. These systems have the same pure states as quantum systems, but different measurement rules, and hence could distinguish ensembles of states which are currently not distinguishable within quantum theory~\cite{mielnik1974, galley2017classification}. The second possibility is that the systems $\A$ and $\B$ are post-quantum systems which are approximated by quantum theory, for instance if the quantum state space is the limit of some discrete state space~\cite{Buniy_2005,Muller_2009, palmer2020discretisation}. This possibility has been argued to arise as a consequence of a discrete space-time~\cite{Buniy_2005,Muller_2009}, which is expected at the interface between quantum theory and gravity.

Concerning (ii) and specifically the possibility for gravity to be neither classical nor quantum, to the best of our knowledge there are three GPTs which allow for entanglement between quantum systems to be generated via non-quantum mediators; that is Refs.~\cite{barnum2016composites}, \cite{hefford2020hyper} and \cite{cavalcanti2021}. The physical realisation of these models is unknown, however it is instructive to consider them as logical possibilities which motivate seeking a theory-independent method to characterise features of the gravitational field. 

This shows that additional conditions are needed to conclude that $\G$ is quantum. Such conditions could shed light on the basic structure of a physical theory at the interface between quantum theory and general relativity.

We remark that assigning a state to the gravitational field, even when the experiment is described exactly in the Newtonian regime of gravity, is supported by arguments that are valid in quantum theory~\cite{belenchia2018quantum, belenchia2019information} (see Appendix~\ref{sec:GravityMed} for a detailed explanation). In these works, failing to assign a quantum state to the Newtonian potential leads to faster-than-light signalling. In a theory-independent discussion like the one presented here, these argument valid in quantum theory are not sufficient to fully justify why the gravitational field should be assigned a GPT state. However, we can use the theory dependent arguments of Appendix~\ref{sec:GravityMed} to further strengthen and motivate the description of the gravitational field $\G$ as a separate GPT system (which may be quantum, classical or some other GPT).

\section{Alternative models of gravity}\label{Sec:ClassGrav}
We now show how our no-go theorem allows us to  classify existing alternative models for the interaction between gravity and matter. For concreteness, we focus on Collapse Models~\cite{ghirardi1986unified, ghirardi1990Markov, Adler_2007, Adler275, bassi2003dynamical}, Gravitational Decoherence~\cite{Anastopoulos:2013zya, blencowe2013effective, bassi2017gravitational}, and the Schr{\"o}dinger-Newton equation~\cite{diosi1984gravitation, diosi1987universal, diosi1989models, penrose1996gravity, penrose2014gravitization, vanMeter:2011xr, anastopoulos2014problems, bahrami2015gravity}.

\begin{table}[]
\centering
	\begin{tabular}{|c|c|c|c|}
	\hline
		& \makecell{Gravitational\\ Decoherence\\ \& 
		Collapse \\models} & \makecell{GPT model \\ corresponding to \\Schr{\"o}dinger-Newton}  & \makecell{Non-classical\\ $\G$}  \\ \hline
		Condition 1 &      \xmark                     &                                                                      \cmark &                                       \cmark          \\ \hline
		Condition 2 &     \cmark                      &                                                                \xmark &                                  \cmark               \\ \hline
		Condition 3 &          \cmark                 &                                                                     \cmark &                                                 \xmark \\ \hline
	\end{tabular}
\caption{Summary of how different proposals violate the different conditions of the no-go theorem. Observe that one could also violate more than one condition. For instance in the case of non-classical $\G$ it could be the case that the field is not a mediator (it acts on both $\A$ and $\B$ globally), and hence condition $2$ is violated also. We note that in gravitational decoherence, although $\G$ is described using quantum theory it is effectively (i.e., operationally) classical.}
\end{table}

\noindent \textbf{Collapse Models} are non-linear modifications to the Schr\"odinger equation which are, however, linear open quantum system dynamics in the density matrix. The relation between Collapse Models and a fundamental theory of gravity is still unknown~\cite{bahrami2015gravity}, however they have been often referred to as accounting for a classical description of gravity. In any case, they can be straightforwardly considered in our no-go theorem, thanks to the fact that they are a quantum system at the level of the density matrix with additional conditions on the dynamics. Specifically, the dynamics leads to a spontaneous localisation of the wavefunction in position space, which prevents the creation of a quantum superposition state of a massive system, and hence of entanglement. They then violate condition 1~\cite{Marletto_2018}.

\noindent \textbf{Gravitational Decoherence} is an effective field theory approach to the quantisation of gravity, in which decoherence arises due to background gravitons. Such a decoherence mechanism ensures that the gravitational field is operationally classical, even though the field is described as a quantum system at the fundamental level. As a consequence, gravity does not lead to entanglement generation in the regime considered, and hence violates condition 1.
	
\noindent\textbf{The Reginatto-Hall model} is a model of interacting quantum and classical ensembles which is claimed to allow for a classical system to generate entanglement between two quantum systems~\cite{reginatto2009quantum, hall2016ensembles}. The model as currently formulated is not a GPT: for instance reduced descriptions of subsystems do not allow one to compute statistics of all future measurements~\cite{hall_communication}, unlike GPT reduced states which are defined to encode all statistics of future measurements. Moreover when interactions are turned off the Reginatto-Hall model exhibits non-local signalling between subsystems, whereas in GPTs subsystems are defined to be non-signalling. These features do not mean there is no GPT associated to the Reginatto-Hall model (i.e. the GPT generated from its operational statistics) but rather that the associated GPT would not share the same subsystem structure. As such the no-go theorem does not apply directly to the Reginatto Hall model itself, but rather states that the GPT associated to the Reginatto-Hall model must violate one of the conditions.

\noindent \textbf{The Schr{\"o}dinger-Newton equation} is usually associated to \emph{semi-classical gravity}, at least under some conditions (e.g., the mean-field approximation)~\cite{bahrami2015gravity, anastopoulos2014problems}. Differently to Collapse Models, nonlinearities are present in the dynamical evolution also at the level of the density matrix. It was shown in Ref.~\cite{mielnik_mobility_1980} that a single-particle system evolving with the Schr{\"o}dinger-Newton equation has a classical state space (i.e., it is a simplex). When two such Schr{\"o}dinger-Newton systems interact gravitationally, we would need to rederive the full set of states and measurements compatible with the dynamical evolution. This is an extremely challenging task, which is beyond the scope of this work. However, we can require some \emph{desiderata} that a GPT corresponding to the Schr{\"o}dinger-Newton equation should satisfy, such as that i) individual systems have a classical state space (i.e., a simplex); ii) composite systems have entangled states (see, e.g., Refs.~\cite{mauro2020classicality, d2020classical} for known examples of classical systems with entanglement). It is usually claimed that nonlinear dynamics, such as the Schr{\"o}dinger-Newton equation, together with the Born rule imply superluminal signalling~\cite{simon2001no}. The proof, however, assumes that mixed states are given by density operators, quantum systems compose according to the standard tensor product rule and that reduced states of a subsystem are obtained via the partial trace. However, this is in general not verified in such nonlinear models. Hence, we cannot conclude that these models violate no-signalling~\cite{masanes_communication}. 

Based on these considerations, a GPT corresponding to the Schr{\"o}dinger-Newton equation violates condition 2 of our no-go theorem. In this scenario, subsystems $\A$ and $\B$ are classical systems with entanglement, so that the composite system $\A\B$ has entangled states; however system $\G$ is classical according to our Definition~\ref{def:classical}, i.e., composite systems $\A\G$ and $\B\G$ do not have entangled states. In this case entanglement generated between $\A$ and $\B$ cannot be mediated by $\G$. Therefore, in this setup conditions 1 and 3 are met while condition 2 is violated. This corresponds, in the Schr{\"o}dinger-Newton equation, to the joint state of $\A$ and $\B$ depending on the field configuration.\\

\section{Discussion}
In this paper, we have introduced the apparatus of Generalised Probabilistic Theories (GPTs) to the study of the nature of the gravitational field. While previous results mostly adopted a specific model of gravity, GPTs allow us to take a model-independent approach and derive general results on the nature of the gravitational field. The long-term goal of this study is to be able to constrain physical theories which are compatible with experimentally testable scenarios, and to systematically check the internal consistency of proposed theories of gravity.	

Here, we have set out the groundwork for achieving this long-term goal by proving a no-go theorem on the nature of the gravitational field, stating that, if gravity-induced entanglement between two masses is detected in an interferometric experiment, then either gravity is not a mediator of the interaction or it has to be non-classical. We have also shown that non-classicality of gravity does not necessarily imply that it is quantum, and we have identified alternative models of systems which are neither classical nor quantum but that can nonetheless entangle quantum systems. 	The theorem also applies when the systems $\A$ and $\B$ which become entangled are not quantum; we give specific examples of cases where they are post-quantum systems, namely they approximate quantum systems in some limit. This shows that our theorem provides a condition for the non-classical nature of gravity which is robust to future modifications to quantum theory. 

In addition, we show that our result also applies to alternative models of gravity such as the Schr{\"o}dinger-Newton equation and Collapse Models, as well as to Gravitational Decoherence.

This work shows how to approach the study of the nature of the gravitational field in a systematic and model independent manner. This may lead to the identification of new fundamental principles which could guide us towards the formulation of a theory unifying general relativity and quantum theory.
A key aim of future work is to provide a full characterisation of all non-classical gravitational theories which can mediate entanglement between quantum systems. This would shed light on the possible features of a unified account of general relativity and quantum theory.\\

\noindent\textbf{Acknowledgments---} The authors would like to thank Markus Aspelmeyer and Markus M{\"u}ller for helpful comments on the first version of the draft; Llu{\'i}s Masanes for pointing out the fact that existing proofs of signalling for NLSE's made an incorrect assumption; Michael Hall for helpful discussions which clarified aspects of the Reginatto-Hall model. T.D.G. and F.G. acknowledge support from Perimeter Institute for Theoretical Physics. Research at Perimeter Institute is supported in part by the Government of Canada through the Department of Innovation, Science and Economic Development and by the Province of Ontario through the Ministry of Colleges and Universities. J.H.S. was supported by the Foundation for Polish Science (IRAP project, ICTQT, contract no. MAB/2018/5, co-financed by EU within Smart Growth Operational Programme). Most of the diagrams were prepared using TikZit.

\appendix

\section{Proof of Theorem~1 (no-go theorem)}\label{proof:ClassIntTwoSys}

For this proof we will work with the diagrammatic representation of GPTs, as this makes the proof much more straightforward than alternative representations. See \cite{chiribella2010probabilistic,hardy2011reformulating,chiribella2011informational,chiribella2016quantum} for details of this formalism. For a very brief primer, however, note that these diagrams should be read bottom to top, GPT systems are denoted by labelled wires, and GPT transformations by labelled boxes. If a box has no ingoing wire then it represents a state of the system and if it has no outgoing wire it represents an effect. We denote discard-preserving transformations (i.e., trace-preserving in the case of quantum theory) by white boxes, and discard-nonincreasing transformations (i.e., trace-nonincreasing transformations in the case of quantum theory) by black boxes. This means that, in particular, we will denote the resolution of the classical identity transformation as:
\beq\label{eq:ROI}
\begin{tikzpicture}
	\begin{pgfonlayer}{nodelayer}
		\node [style=none] (31) at (3, -1.5) {};
		\node [style=none] (32) at (3, 1.5) {};
		\node [style=right label] (33) at (3, -1) {$X$};
	\end{pgfonlayer}
	\begin{pgfonlayer}{edgelayer}
		\draw (31.center) to (32.center);
	\end{pgfonlayer}
\end{tikzpicture}
\quad = \quad \begin{tikzpicture}
	\begin{pgfonlayer}{nodelayer}
		\node [style=copoint, fill=black] (16) at (3, -1) {$\color{white}x$};
		\node [style=none] (17) at (3, -2) {};
		\node [style=right label] (18) at (3, -1.75) {$X$};
		\node [style=point] (31) at (3, 0.5) {$x$};
		\node [style=none] (32) at (3, 1.5) {};
		\node [style=right label] (33) at (3, 1) {$X$};
		\node [style=none] (34) at (1.5, -0.25) {$\displaystyle\sum_{x\in X}$};
	\end{pgfonlayer}
	\begin{pgfonlayer}{edgelayer}
		\draw (16) to (17.center);
		\draw (31) to (32.center);
	\end{pgfonlayer}
\end{tikzpicture}.
\eeq
In this diagram the states
\beq
\begin{tikzpicture}
	\begin{pgfonlayer}{nodelayer}
		\node [style=point] (3) at (0, -0.25) {$x$};
		\node [style=none] (4) at (0, 0.75) {};
		\node [style=right label] (5) at (0, 0.25) {$X$};
	\end{pgfonlayer}
	\begin{pgfonlayer}{edgelayer}
		\draw (3) to (4.center);
	\end{pgfonlayer}
\end{tikzpicture}
\eeq
correspond to the delta function distributions on the sample-space $X$ and the effects
\beq
\begin{tikzpicture}
	\begin{pgfonlayer}{nodelayer}
		\node [style=copoint, fill=black] (0) at (5, 0.25) {$\color{white}x$};
		\node [style=none] (1) at (5, -0.75) {};
		\node [style=right label] (2) at (5, -0.5) {$X$};
	\end{pgfonlayer}
	\begin{pgfonlayer}{edgelayer}
		\draw (0) to (1.center);
	\end{pgfonlayer}
\end{tikzpicture}
\eeq
are the effects that perfectly distinguish between the different elements of the sample-space, that is, they satisfy:
\beq
\begin{tikzpicture}
	\begin{pgfonlayer}{nodelayer}
		\node [style=copoint, fill=black] (0) at (0, 0.75) {$\color{white}x'$};
		\node [style=none] (1) at (0, 0) {};
		\node [style=point] (3) at (0, -0.75) {$x$};
		\node [style=none] (4) at (0, 0) {};
		\node [style=right label] (5) at (0, 0) {$X$};
	\end{pgfonlayer}
	\begin{pgfonlayer}{edgelayer}
		\draw (0) to (1.center);
		\draw (3) to (4.center);
	\end{pgfonlayer}
\end{tikzpicture}
\quad = \ \ \delta_{x,x'}.
\eeq

Note that the underlying operational structure (i.e., symmetric monoidal categorical structure) of the GPT means that diagrams are invariant under moving transformations around on the page so long as the way they are connected up is preserved, whilst the probabilistic structure underpinning the GPT framework ensures that the sum in Eq.~\eqref{eq:ROI} is also free to be moved around the diagram.

Let us observe that the statement of Theorem~1 is equivalent to saying that a classical system cannot mediate entanglement via local interactions. 

Therefore we can rephrase our no-go theorem in the following manner:

\begin{theorem}[Rephrased] \label{Th:ClassIntTwoSys}
In any GPT, interactions between any two systems, $\A$ and $\mathsf{B}$, which are mediated by classical systems, $\G$, cannot generate entanglement.
\end{theorem}
\proof
Let us first start with a simple example as the full proof is a straightforward generalisation thereof. An example of an interaction between $\A$ and $\B$ mediated by $\G$ is given by the following diagram:
\beq
\begin{tikzpicture}
	\begin{pgfonlayer}{nodelayer}
		\node [style=none] (0) at (-2, -1.75) {};
		\node [style=none] (1) at (-2, 0) {};
		\node [style=right label] (2) at (-2, -1.5) {$\A$};
		\node [style=none] (3) at (2, -1.75) {};
		\node [style=none] (4) at (2, 2) {};
		\node [style=right label] (5) at (2, -1.5) {$\B$};
		\node [style=none] (6) at (0, 0) {};
		\node [style=point] (7) at (0, -1) {$s$};
		\node [style=right label] (8) at (0, -0.5) {$\G$};
		\node [style=none] (9) at (-2.5, 0) {};
		\node [style=none] (10) at (-2.5, 1) {};
		\node [style=none] (11) at (0.5, 1) {};
		\node [style=none] (12) at (0.5, 0) {};
		\node [style=none] (13) at (-2, 1) {};
		\node [style=none] (14) at (-2, 4) {};
		\node [style=right label] (15) at (-2, 3.5) {$\A$};
		\node [style=none] (16) at (0, 2) {};
		\node [style=none] (17) at (0, 1) {};
		\node [style=right label] (18) at (0, 1.25) {$\G$};
		\node [style=none] (21) at (-0.5, 2) {};
		\node [style=none] (22) at (-0.5, 3) {};
		\node [style=none] (23) at (2.5, 3) {};
		\node [style=none] (24) at (2.5, 2) {};
		\node [style=none] (25) at (0, 3) {};
		\node [style=none] (26) at (0, 4) {};
		\node [style=none] (28) at (2, 4) {};
		\node [style=none] (29) at (2, 3) {};
		\node [style=right label] (30) at (0, 3.5) {$\G$};
		\node [style=right label] (31) at (2, 3.5) {$\B$};
		\node [style=none] (32) at (-1, 0.5) {$\mathcal{I}_A$};
		\node [style=none] (33) at (1, 2.5) {$\mathcal{I}_B$};
	\end{pgfonlayer}
	\begin{pgfonlayer}{edgelayer}
		\draw [qWire] (1.center) to (0.center);
		\draw [qWire] (4.center) to (3.center);
		\draw (6.center) to (7);
		\draw (10.center) to (11.center);
		\draw (11.center) to (12.center);
		\draw (12.center) to (9.center);
		\draw (9.center) to (10.center);
		\draw [qWire] (14.center) to (13.center);
		\draw (16.center) to (17.center);
		\draw (22.center) to (23.center);
		\draw (23.center) to (24.center);
		\draw (24.center) to (21.center);
		\draw (21.center) to (22.center);
		\draw (26.center) to (25.center);
		\draw [qWire] (28.center) to (29.center);
	\end{pgfonlayer}
\end{tikzpicture}
\eeq
That is, we have the initial state of the field $s$, which then first interacts with the system $\A$ via the interaction $\mathcal{I}_A$ and then with the system $\B$ via the interaction $\mathcal{I}_B$.

To see that this is incapable of generating entanglement, we use the fact that $\G$ is classical in order to insert a resolution of the identity in between the two interactions, and post select on some final field state $s'$:
\beq
\begin{tikzpicture}
	\begin{pgfonlayer}{nodelayer}
		\node [style=none] (0) at (-2, -4.75) {};
		\node [style=none] (1) at (-2, -3) {};
		\node [style=right label] (2) at (-2, -4.5) {$\A$};
		\node [style=none] (3) at (2, -4.5) {};
		\node [style=none] (4) at (2, 1.5) {};
		\node [style=right label] (5) at (2, -4.25) {$\B$};
		\node [style=none] (6) at (0, -3) {};
		\node [style=point] (7) at (0, -4) {$s$};
		\node [style=right label] (8) at (0, -3.5) {$\G$};
		\node [style=none] (9) at (-2.5, -3) {};
		\node [style=none] (10) at (-2.5, -2) {};
		\node [style=none] (11) at (0.5, -2) {};
		\node [style=none] (12) at (0.5, -3) {};
		\node [style=none] (13) at (-2, -2) {};
		\node [style=none] (14) at (-2, 4.25) {};
		\node [style=right label] (15) at (-2, 3.75) {$\A$};
		\node [style=copoint, fill=black] (16) at (0, -1) {$\color{white}f$};
		\node [style=none] (17) at (0, -2) {};
		\node [style=right label] (18) at (0, -1.75) {$\G$};
		\node [style=none] (21) at (-0.5, 1.5) {};
		\node [style=none] (22) at (-0.5, 2.5) {};
		\node [style=none] (23) at (2.5, 2.5) {};
		\node [style=none] (24) at (2.5, 1.5) {};
		\node [style=none] (25) at (0, 2.5) {};
		\node [style=copoint, fill=black] (26) at (0, 3.5) {$\color{white}s'$};
		\node [style=none] (28) at (2, 4.25) {};
		\node [style=none] (29) at (2, 2.5) {};
		\node [style=right label] (30) at (0, 2.75) {$\G$};
		\node [style=right label] (31) at (2, 3.75) {$\B$};
		\node [style=none] (32) at (-1, -2.5) {$\mathcal{I}_A$};
		\node [style=none] (33) at (1, 2) {$\mathcal{I}_B$};
		\node [style=point] (34) at (0, 0.5) {$f$};
		\node [style=none] (35) at (0, 1.5) {};
		\node [style=right label] (36) at (0, 1) {$\G$};
		\node [style=none] (37) at (-1, -0.25) {$\displaystyle\sum_f$};
	\end{pgfonlayer}
	\begin{pgfonlayer}{edgelayer}
		\draw [qWire] (1.center) to (0.center);
		\draw [qWire] (4.center) to (3.center);
		\draw (6.center) to (7);
		\draw (10.center) to (11.center);
		\draw (11.center) to (12.center);
		\draw (12.center) to (9.center);
		\draw (9.center) to (10.center);
		\draw [qWire] (14.center) to (13.center);
		\draw (16) to (17.center);
		\draw (22.center) to (23.center);
		\draw (23.center) to (24.center);
		\draw (24.center) to (21.center);
		\draw (21.center) to (22.center);
		\draw (26) to (25.center);
		\draw [qWire] (28.center) to (29.center);
		\draw (34) to (35.center);
	\end{pgfonlayer}
\end{tikzpicture}
\eeq
Using simple diagrammatic manipulations this can be rewritten as follows:
\beq\label{eq:proof1}\sum_f
\begin{tikzpicture}
	\begin{pgfonlayer}{nodelayer}
		\node [style=none] (0) at (-4, -4.25) {};
		\node [style=none] (1) at (-4, -0.5) {};
		\node [style=right label] (2) at (-4, -4) {$\A$};
		\node [style=none] (3) at (2, -4.5) {};
		\node [style=none] (4) at (2, -0.5) {};
		\node [style=right label] (5) at (2, -4.25) {$\B$};
		\node [style=none] (6) at (-2, -0.5) {};
		\node [style=point] (7) at (-2, -1.5) {$s$};
		\node [style=right label] (8) at (-2, -1) {$\G$};
		\node [style=none] (9) at (-4.5, -0.5) {};
		\node [style=none] (10) at (-4.5, 0.5) {};
		\node [style=none] (11) at (-1.5, 0.5) {};
		\node [style=none] (12) at (-1.5, -0.5) {};
		\node [style=none] (13) at (-4, 0.5) {};
		\node [style=none] (14) at (-4, 4.25) {};
		\node [style=right label] (15) at (-4, 3.75) {$\A$};
		\node [style=copoint,fill=black] (16) at (-2, 1.5) {$\color{white}f$};
		\node [style=none] (17) at (-2, 0.5) {};
		\node [style=right label] (18) at (-2, 0.75) {$\G$};
		\node [style=none] (21) at (-0.5, -0.5) {};
		\node [style=none] (22) at (-0.5, 0.5) {};
		\node [style=none] (23) at (2.5, 0.5) {};
		\node [style=none] (24) at (2.5, -0.5) {};
		\node [style=none] (25) at (0, 0.5) {};
		\node [style=copoint,fill=black] (26) at (0, 1.5) {$\color{white}s'$};
		\node [style=none] (28) at (2, 4.25) {};
		\node [style=none] (29) at (2, 0.5) {};
		\node [style=right label] (30) at (0, .75) {$\G$};
		\node [style=right label] (31) at (2, 3.75) {$\B$};
		\node [style=none] (32) at (-3, 0) {$\mathcal{I}_A$};
		\node [style=none] (33) at (1, 0) {$\mathcal{I}_B$};
		\node [style=point] (34) at (0, -1.5) {$f$};
		\node [style=none] (35) at (0, -0.5) {};
		\node [style=right label] (36) at (0, -1) {$\G$};
		\node [style=none] (37) at (-5, 2.5) {};
		\node [style=none] (38) at (-1, 2.5) {};
		\node [style=none] (39) at (-1, -2.5) {};
		\node [style=none] (40) at (-5, -2.5) {};
		\node [style=none] (41) at (-0.75, 2.5) {};
		\node [style=none] (42) at (3, 2.5) {};
		\node [style=none] (43) at (3, -2.5) {};
		\node [style=none] (44) at (-0.75, -2.5) {};
	\end{pgfonlayer}
	\begin{pgfonlayer}{edgelayer}
		\draw [qWire] (1.center) to (0.center);
		\draw [qWire] (4.center) to (3.center);
		\draw (6.center) to (7);
		\draw (10.center) to (11.center);
		\draw (11.center) to (12.center);
		\draw (12.center) to (9.center);
		\draw (9.center) to (10.center);
		\draw [qWire] (14.center) to (13.center);
		\draw (16) to (17.center);
		\draw (22.center) to (23.center);
		\draw (23.center) to (24.center);
		\draw (24.center) to (21.center);
		\draw (21.center) to (22.center);
		\draw (26) to (25.center);
		\draw [qWire] (28.center) to (29.center);
		\draw (34) to (35.center);
		\draw  [thick gray dashed edge](37.center) to (38.center);
		\draw  [thick gray dashed edge](38.center) to (39.center);
		\draw [thick gray dashed edge] (39.center) to (40.center);
		\draw  [thick gray dashed edge](40.center) to (37.center);
		\draw  [thick gray dashed edge](41.center) to (42.center);
		\draw  [thick gray dashed edge](42.center) to (43.center);
		\draw  [thick gray dashed edge](43.center) to (44.center);
		\draw [thick gray dashed edge](44.center) to (41.center);
	\end{pgfonlayer}
\end{tikzpicture}
\ \ =: \ \sum_f \begin{tikzpicture}
	\begin{pgfonlayer}{nodelayer}
		\node [style=none] (0) at (1.25, -2.25) {};
		\node [style=none] (1) at (1.25, -0.75) {};
		\node [style=right label] (2) at (1.25, -2) {$\A$};
		\node [style=none] (3) at (3.25, -2.25) {};
		\node [style=none] (4) at (3.25, -0.75) {};
		\node [style=right label] (5) at (3.25, -2) {$\B$};
		\node [style=none] (6) at (0.5, -0.75) {};
		\node [style=none] (7) at (0.5, 0.75) {};
		\node [style=none] (8) at (2, 0.75) {};
		\node [style=none] (9) at (2, -0.75) {};
		\node [style=none] (10) at (1.25, 0.75) {};
		\node [style=none] (11) at (1.25, 2.25) {};
		\node [style=right label] (12) at (1.25, 1.75) {$\A$};
		\node [style=none] (13) at (2.5, -0.75) {};
		\node [style=none] (14) at (2.5, 0.75) {};
		\node [style=none] (15) at (4, 0.75) {};
		\node [style=none] (16) at (4, -0.75) {};
		\node [style=none] (17) at (3.25, 2.25) {};
		\node [style=none] (18) at (3.25, 0.75) {};
		\node [style=right label] (19) at (3.25, 1.75) {$\B$};
		\node [style=none] (20) at (1.25, 0) {$\color{white}\mathcal{E}_A^f$};
		\node [style=none] (21) at (3.25, 0) {$\color{white}\mathcal{E}_B^f$};
	\end{pgfonlayer}
	\begin{pgfonlayer}{edgelayer}
		\draw [qWire] (1.center) to (0.center);
		\draw [qWire] (4.center) to (3.center);
		\draw [fill=black] (9.center)
			 to (6.center)
			 to (7.center)
			 to (8.center)
			 to cycle;
		\draw [qWire] (11.center) to (10.center);
		\draw [fill=black] (16.center)
			 to (13.center)
			 to (14.center)
			 to (15.center)
			 to cycle;
		\draw [qWire] (17.center) to (18.center);
	\end{pgfonlayer}
\end{tikzpicture}
.
\eeq
It is clear that $\mathcal{E}_A^f$ and $\mathcal{E}_B^f$ are all discard-nonincreasing transformations (for $\A$ and $\B$ respectively) as they are composite of other discard-nonincreasing transformations.

It is then not hard to see that such sums of product discard-nonincreasing transformations are incapable of generating entanglement. A separable state $S$, can be written as:
\beq
\begin{tikzpicture}
	\begin{pgfonlayer}{nodelayer}
		\node [style=none] (0) at (0, 0) {$S$};
		\node [style=none] (1) at (-1.25, 0.5) {};
		\node [style=none] (2) at (1.25, 0.5) {};
		\node [style=none] (3) at (0, -0.75) {};
		\node [style=none] (4) at (-0.75, 0.5) {};
		\node [style=none] (5) at (-0.75, 1.25) {};
		\node [style=right label] (6) at (-0.75, 1) {$\A$};
		\node [style=none] (7) at (0.75, 0.5) {};
		\node [style=none] (8) at (0.75, 1.25) {};
		\node [style=right label] (9) at (0.75, 1) {$\B$};
	\end{pgfonlayer}
	\begin{pgfonlayer}{edgelayer}
		\draw (3.center) to (1.center);
		\draw (2.center) to (1.center);
		\draw (3.center) to (2.center);
		\draw [qWire] (5.center) to (4.center);
		\draw [qWire] (8.center) to (7.center);
	\end{pgfonlayer}
\end{tikzpicture}
\quad = \quad \sum_i \begin{tikzpicture}
	\begin{pgfonlayer}{nodelayer}
		\node [style=point,fill=black] (4) at (-0.75, 0) {\color{white}$s^A_i$};
		\node [style=none] (5) at (-0.75, 1.25) {};
		\node [style=right label] (6) at (-0.75, 1) {$\A$};
		\node [style=point,fill=black] (7) at (0.75, 0) {\color{white}$s^B_i$};
		\node [style=none] (8) at (0.75, 1.25) {};
		\node [style=right label] (9) at (0.75, 1) {$\B$};
	\end{pgfonlayer}
	\begin{pgfonlayer}{edgelayer}
		\draw [qWire] (5.center) to (4);
		\draw [qWire] (8.center) to (7);
	\end{pgfonlayer}
\end{tikzpicture}
\eeq
where the $s^A_i$ and $s^B_i$ are (possibly subnormalised states of $\A$ and $\B$ respectively. If we compose a separable state with a transformation of the form of Eq.~\eqref{eq:proof1} we find that we obtain another separable state:
\beq
\sum_{i,f}  \begin{tikzpicture}
	\begin{pgfonlayer}{nodelayer}
		\node [style=point, fill=black] (0) at (1.25, -1.75) {\color{white}$s^A_i$};
		\node [style=none] (1) at (1.25, -0.5) {};
		\node [style=right label] (2) at (1.25, -1) {$\A$};
		\node [style=point, fill=black] (3) at (3.25, -1.75) {\color{white}$s^B_i$};
		\node [style=none] (4) at (3.25, -0.5) {};
		\node [style=right label] (5) at (3.25, -1) {$\B$};
		\node [style=none] (6) at (0.5, -0.5) {};
		\node [style=none] (7) at (0.5, 1) {};
		\node [style=none] (8) at (2, 1) {};
		\node [style=none] (9) at (2, -0.5) {};
		\node [style=none] (10) at (1.25, 1) {};
		\node [style=none] (11) at (1.25, 2.5) {};
		\node [style=right label] (12) at (1.25, 2) {$\A$};
		\node [style=none] (13) at (2.5, -0.5) {};
		\node [style=none] (14) at (2.5, 1) {};
		\node [style=none] (15) at (4, 1) {};
		\node [style=none] (16) at (4, -0.5) {};
		\node [style=none] (17) at (3.25, 2.5) {};
		\node [style=none] (18) at (3.25, 1) {};
		\node [style=right label] (19) at (3.25, 2) {$\B$};
		\node [style=none] (20) at (1.25, 0.25) {$\color{white}\mathcal{E}_A^f$};
		\node [style=none] (21) at (3.25, 0.25) {$\color{white}\mathcal{E}_B^f$};
	\end{pgfonlayer}
	\begin{pgfonlayer}{edgelayer}
		\draw [qWire] (1.center) to (0);
		\draw [qWire] (4.center) to (3);
		\draw [fill=black] (9.center)
			 to (6.center)
			 to (7.center)
			 to (8.center)
			 to cycle;
		\draw [qWire] (11.center) to (10.center);
		\draw [fill=black] (16.center)
			 to (13.center)
			 to (14.center)
			 to (15.center)
			 to cycle;
		\draw [qWire] (17.center) to (18.center);
	\end{pgfonlayer}
\end{tikzpicture}
\quad =: \quad \sum_{i,f} \begin{tikzpicture}
	\begin{pgfonlayer}{nodelayer}
		\node [style=point, fill=black] (0) at (0, 0) {\color{white}${s'}^A_{i,f}$};
		\node [style=none] (1) at (0, 1.25) {};
		\node [style=right label] (2) at (0, 1) {$\A$};
		\node [style=point, fill=black] (3) at (2.5, 0) {\color{white}${s'}^B_{i,f}$};
		\node [style=none] (4) at (2.5, 1.25) {};
		\node [style=right label] (5) at (2.5, 1) {$\B$};
	\end{pgfonlayer}
	\begin{pgfonlayer}{edgelayer}
		\draw [qWire] (1.center) to (0);
		\draw [qWire] (4.center) to (3);
	\end{pgfonlayer}
\end{tikzpicture}
.
\eeq

This, however, is just a simple example of an interaction between two systems mediated by a classical gravitational field. One may therefore be tempted to speculate that some more complicated interaction could generate entanglement.

The most general form that such an interaction can take can be written as a sequence of pairwise interactions of the above form, for example:
\begin{align}
\begin{tikzpicture}
	\begin{pgfonlayer}{nodelayer}
		\node [style=none] (0) at (-2.5, -4) {};
		\node [style=none] (1) at (-2.5, -3.25) {};
		\node [style=right label] (2) at (-2.5, -3.75) {$\A$};
		\node [style=none] (3) at (2.5, -4) {};
		\node [style=none] (4) at (2.5, -3.25) {};
		\node [style=right label] (5) at (2.5, -3.75) {$\B$};
		\node [style=none] (6) at (-1, -3.25) {};
		\node [style=none] (7) at (-1, -4) {};
		\node [style=right label] (8) at (-1, -3.75) {$\G_A$};
		\node [style=none] (9) at (-3, -3.25) {};
		\node [style=none] (10) at (-3, -2.25) {};
		\node [style=none] (11) at (-0.5, -2.25) {};
		\node [style=none] (12) at (-0.5, -3.25) {};
		\node [style=none] (13) at (-2.5, -2.25) {};
		\node [style=none] (14) at (-2.5, 0.25) {};
		\node [style=right label] (15) at (-2.5, -2) {$\A$};
		\node [style=none] (19) at (0.5, -3.25) {};
		\node [style=none] (20) at (0.5, -2.25) {};
		\node [style=none] (21) at (3, -2.25) {};
		\node [style=none] (22) at (3, -3.25) {};
		\node [style=none] (25) at (2.5, 0.25) {};
		\node [style=none] (26) at (2.5, -2.25) {};
		\node [style=right label] (28) at (2.5, -2) {$\B$};
		\node [style=none] (31) at (-2.5, 0.25) {};
		\node [style=none] (75) at (0, -1.5) {};
		\node [style=none] (76) at (0, -4) {};
		\node [style=right label] (77) at (0, -3.75) {$\G_R$};
		\node [style=none] (78) at (1, -3.25) {};
		\node [style=none] (79) at (1, -4) {};
		\node [style=right label] (80) at (1, -3.75) {$\G_B$};
		\node [style=none] (81) at (-1, -1.5) {};
		\node [style=none] (82) at (-1, -2.25) {};
		\node [style=right label] (83) at (-1, -2) {$\G_A$};
		\node [style=none] (84) at (1, -1.5) {};
		\node [style=none] (85) at (1, -2.25) {};
		\node [style=right label] (86) at (1, -2) {$\G_B$};
		\node [style=none] (88) at (-1.5, -1.5) {};
		\node [style=none] (89) at (-1.5, -0.5) {};
		\node [style=none] (90) at (1.5, -0.5) {};
		\node [style=none] (91) at (1.5, -1.5) {};
		\node [style=none] (92) at (-1, 0.25) {};
		\node [style=none] (93) at (-1, -0.5) {};
		\node [style=right label] (94) at (-1, -0.25) {$\G_A$};
		\node [style=none] (99) at (0, 2) {};
		\node [style=none] (100) at (0, -0.5) {};
		\node [style=right label] (101) at (0, -0.25) {$\G_R$};
		\node [style=none] (102) at (1, 0.25) {};
		\node [style=none] (103) at (1, -0.5) {};
		\node [style=right label] (104) at (1, -0.25) {$\G_B$};
		\node [style=none] (114) at (-3, 0.25) {};
		\node [style=none] (115) at (-3, 1.25) {};
		\node [style=none] (116) at (-0.5, 1.25) {};
		\node [style=none] (117) at (-0.5, 0.25) {};
		\node [style=none] (118) at (-2.5, 1.25) {};
		\node [style=none] (119) at (-2.5, 3.75) {};
		\node [style=right label] (120) at (-2.5, 1.5) {$\A$};
		\node [style=none] (121) at (0.5, 0.25) {};
		\node [style=none] (122) at (0.5, 1.25) {};
		\node [style=none] (123) at (3, 1.25) {};
		\node [style=none] (124) at (3, 0.25) {};
		\node [style=none] (125) at (2.5, 3.75) {};
		\node [style=none] (126) at (2.5, 1.25) {};
		\node [style=right label] (127) at (2.5, 1.5) {$\B$};
		\node [style=none] (128) at (-2.5, 3.75) {};
		\node [style=none] (136) at (-1, 2) {};
		\node [style=none] (137) at (-1, 1.25) {};
		\node [style=right label] (138) at (-1, 1.5) {$\G_A$};
		\node [style=none] (139) at (1, 2) {};
		\node [style=none] (140) at (1, 1.25) {};
		\node [style=right label] (141) at (1, 1.5) {$\G_B$};
		\node [style=none] (142) at (-1.5, 2) {};
		\node [style=none] (143) at (-1.5, 3) {};
		\node [style=none] (144) at (1.5, 3) {};
		\node [style=none] (145) at (1.5, 2) {};
		\node [style=none] (146) at (-1, 3.75) {};
		\node [style=none] (147) at (-1, 3) {};
		\node [style=right label] (148) at (-1, 3.25) {$\G_A$};
		\node [style=none] (149) at (0, 3.75) {};
		\node [style=none] (150) at (0, 3) {};
		\node [style=right label] (151) at (0, 3.25) {$\G_R$};
		\node [style=none] (152) at (1, 3.75) {};
		\node [style=none] (153) at (1, 3) {};
		\node [style=right label] (154) at (1, 3.25) {$\G_B$};
	\end{pgfonlayer}
	\begin{pgfonlayer}{edgelayer}
		\draw [qWire] (1.center) to (0.center);
		\draw [qWire] (4.center) to (3.center);
		\draw (6.center) to (7.center);
		\draw (10.center) to (11.center);
		\draw (11.center) to (12.center);
		\draw (12.center) to (9.center);
		\draw (9.center) to (10.center);
		\draw [qWire] (14.center) to (13.center);
		\draw (20.center) to (21.center);
		\draw (21.center) to (22.center);
		\draw (22.center) to (19.center);
		\draw (19.center) to (20.center);
		\draw [qWire] (25.center) to (26.center);
		\draw (75.center) to (76.center);
		\draw (78.center) to (79.center);
		\draw (81.center) to (82.center);
		\draw (84.center) to (85.center);
		\draw (89.center) to (90.center);
		\draw (90.center) to (91.center);
		\draw (91.center) to (88.center);
		\draw (88.center) to (89.center);
		\draw (92.center) to (93.center);
		\draw (99.center) to (100.center);
		\draw (102.center) to (103.center);
		\draw (115.center) to (116.center);
		\draw (116.center) to (117.center);
		\draw (117.center) to (114.center);
		\draw (114.center) to (115.center);
		\draw [qWire] (119.center) to (118.center);
		\draw (122.center) to (123.center);
		\draw (123.center) to (124.center);
		\draw (124.center) to (121.center);
		\draw (121.center) to (122.center);
		\draw [qWire] (125.center) to (126.center);
		\draw (136.center) to (137.center);
		\draw (139.center) to (140.center);
		\draw (143.center) to (144.center);
		\draw (144.center) to (145.center);
		\draw (145.center) to (142.center);
		\draw (142.center) to (143.center);
		\draw (146.center) to (147.center);
		\draw (149.center) to (150.center);
		\draw (152.center) to (153.center);
	\end{pgfonlayer}
\end{tikzpicture}} \ &= \  %
\begin{tikzpicture}
	\begin{pgfonlayer}{nodelayer}
		\node [style=none] (0) at (-2.25, -3.75) {};
		\node [style=none] (1) at (-2.25, -3) {};
		\node [style=right label] (2) at (-2.25, -3.5) {$\A$};
		\node [style=none] (3) at (2.75, -3.75) {};
		\node [style=none] (4) at (2.75, -3) {};
		\node [style=right label] (5) at (2.75, -3.5) {$\B$};
		\node [style=none] (6) at (-0.75, -3) {};
		\node [style=none] (7) at (-0.75, -3.75) {};
		\node [style=right label] (8) at (-0.75, -3.5) {$\G_A$};
		\node [style=none] (9) at (-2.75, -3) {};
		\node [style=none] (10) at (-2.75, -2) {};
		\node [style=none] (11) at (-0.25, -2) {};
		\node [style=none] (12) at (-0.25, -3) {};
		\node [style=none] (13) at (-2.25, -2) {};
		\node [style=none] (14) at (-2.25, 0.5) {};
		\node [style=right label] (15) at (-2.25, -1.75) {$\A$};
		\node [style=none] (19) at (0.75, -3) {};
		\node [style=none] (20) at (0.75, -2) {};
		\node [style=none] (21) at (3.25, -2) {};
		\node [style=none] (22) at (3.25, -3) {};
		\node [style=none] (25) at (2.75, 0.5) {};
		\node [style=none] (26) at (2.75, -2) {};
		\node [style=right label] (28) at (2.75, -1.75) {$\B$};
		\node [style=none] (31) at (-2.25, 0.5) {};
		\node [style=none] (75) at (0.25, -1.25) {};
		\node [style=none] (76) at (0.25, -3.75) {};
		\node [style=right label] (77) at (0.25, -3.5) {$\G_R$};
		\node [style=none] (78) at (1.25, -3) {};
		\node [style=none] (79) at (1.25, -3.75) {};
		\node [style=right label] (80) at (1.25, -3.5) {$\G_B$};
		\node [style=none] (81) at (-0.75, -1.25) {};
		\node [style=none] (82) at (-0.75, -2) {};
		\node [style=right label] (83) at (-0.75, -1.75) {$\G_A$};
		\node [style=none] (84) at (1.25, -1.25) {};
		\node [style=none] (85) at (1.25, -2) {};
		\node [style=right label] (86) at (1.25, -1.75) {$\G_B$};
		\node [style=none] (88) at (-1.25, -1.25) {};
		\node [style=none] (89) at (-1.25, -0.25) {};
		\node [style=none] (90) at (1.75, -0.25) {};
		\node [style=none] (91) at (1.75, -1.25) {};
		\node [style=none] (92) at (-0.75, 0.5) {};
		\node [style=none] (93) at (-0.75, -0.25) {};
		\node [style=right label] (94) at (-0.75, 0) {$\G_A$};
		\node [style=none] (99) at (0.25, 2.25) {};
		\node [style=none] (100) at (0.25, -0.25) {};
		\node [style=right label] (101) at (0.25, 0) {$\G_R$};
		\node [style=none] (102) at (1.25, 0.5) {};
		\node [style=none] (103) at (1.25, -0.25) {};
		\node [style=right label] (104) at (1.25, 0) {$\G_B$};
		\node [style=none] (114) at (-2.75, 0.5) {};
		\node [style=none] (115) at (-2.75, 1.5) {};
		\node [style=none] (116) at (-0.25, 1.5) {};
		\node [style=none] (117) at (-0.25, 0.5) {};
		\node [style=none] (118) at (-2.25, 1.5) {};
		\node [style=none] (119) at (-2.25, 4) {};
		\node [style=right label] (120) at (-2.25, 1.75) {$\A$};
		\node [style=none] (121) at (0.75, 0.5) {};
		\node [style=none] (122) at (0.75, 1.5) {};
		\node [style=none] (123) at (3.25, 1.5) {};
		\node [style=none] (124) at (3.25, 0.5) {};
		\node [style=none] (125) at (2.75, 4) {};
		\node [style=none] (126) at (2.75, 1.5) {};
		\node [style=right label] (127) at (2.75, 1.75) {$\B$};
		\node [style=none] (128) at (-2.25, 4) {};
		\node [style=none] (129) at (2.75, 4) {};
		\node [style=none] (136) at (-0.75, 2.25) {};
		\node [style=none] (137) at (-0.75, 1.5) {};
		\node [style=right label] (138) at (-0.75, 1.75) {$\G_A$};
		\node [style=none] (139) at (1.25, 2.25) {};
		\node [style=none] (140) at (1.25, 1.5) {};
		\node [style=right label] (141) at (1.25, 1.75) {$\G_B$};
		\node [style=none] (142) at (-1.25, 2.25) {};
		\node [style=none] (143) at (-1.25, 3.25) {};
		\node [style=none] (144) at (1.75, 3.25) {};
		\node [style=none] (145) at (1.75, 2.25) {};
		\node [style=none] (146) at (-0.75, 4) {};
		\node [style=none] (147) at (-0.75, 3.25) {};
		\node [style=right label] (148) at (-0.75, 3.5) {$\G_A$};
		\node [style=none] (149) at (0.25, 4) {};
		\node [style=none] (150) at (0.25, 3.25) {};
		\node [style=right label] (151) at (0.25, 3.5) {$\G_R$};
		\node [style=none] (152) at (1.25, 4) {};
		\node [style=none] (153) at (1.25, 3.25) {};
		\node [style=right label] (154) at (1.25, 3.5) {$\G_B$};
		\node [style=none] (155) at (-1.5, 3.5) {};
		\node [style=none] (156) at (-1.5, 2) {};
		\node [style=none] (157) at (0.5, 2) {};
		\node [style=none] (158) at (0.5, 0.25) {};
		\node [style=none] (159) at (3.5, 0.25) {};
		\node [style=none] (160) at (3.5, 1.75) {};
		\node [style=none] (161) at (2, 1.75) {};
		\node [style=none] (162) at (2, 3.5) {};
		\node [style=none] (163) at (-1.5, 0) {};
		\node [style=none] (164) at (-1.5, -1.5) {};
		\node [style=none] (165) at (0.5, -1.5) {};
		\node [style=none] (166) at (0.5, -3.25) {};
		\node [style=none] (167) at (3.5, -3.25) {};
		\node [style=none] (168) at (3.5, -1.75) {};
		\node [style=none] (169) at (2, -1.75) {};
		\node [style=none] (170) at (2, 0) {};
	\end{pgfonlayer}
	\begin{pgfonlayer}{edgelayer}
		\draw [qWire] (1.center) to (0.center);
		\draw [qWire] (4.center) to (3.center);
		\draw (6.center) to (7.center);
		\draw (10.center) to (11.center);
		\draw (11.center) to (12.center);
		\draw (12.center) to (9.center);
		\draw (9.center) to (10.center);
		\draw [qWire] (14.center) to (13.center);
		\draw (20.center) to (21.center);
		\draw (21.center) to (22.center);
		\draw (22.center) to (19.center);
		\draw (19.center) to (20.center);
		\draw [qWire] (25.center) to (26.center);
		\draw (75.center) to (76.center);
		\draw (78.center) to (79.center);
		\draw (81.center) to (82.center);
		\draw (84.center) to (85.center);
		\draw (89.center) to (90.center);
		\draw (90.center) to (91.center);
		\draw (91.center) to (88.center);
		\draw (88.center) to (89.center);
		\draw (92.center) to (93.center);
		\draw (99.center) to (100.center);
		\draw (102.center) to (103.center);
		\draw (115.center) to (116.center);
		\draw (116.center) to (117.center);
		\draw (117.center) to (114.center);
		\draw (114.center) to (115.center);
		\draw [qWire] (119.center) to (118.center);
		\draw (122.center) to (123.center);
		\draw (123.center) to (124.center);
		\draw (124.center) to (121.center);
		\draw (121.center) to (122.center);
		\draw [qWire] (125.center) to (126.center);
		\draw (136.center) to (137.center);
		\draw (139.center) to (140.center);
		\draw (143.center) to (144.center);
		\draw (144.center) to (145.center);
		\draw (145.center) to (142.center);
		\draw (142.center) to (143.center);
		\draw (146.center) to (147.center);
		\draw (149.center) to (150.center);
		\draw (152.center) to (153.center);
		\draw [thick gray dashed edge] (160.center)
			 to (159.center)
			 to (158.center)
			 to (157.center)
			 to (156.center)
			 to (155.center)
			 to (162.center)
			 to (161.center)
			 to cycle;
		\draw [thick gray dashed edge] (168.center)
			 to (167.center)
			 to (166.center)
			 to (165.center)
			 to (164.center)
			 to (163.center)
			 to (170.center)
			 to (169.center)
			 to cycle;
	\end{pgfonlayer}
\end{tikzpicture}} \\
&=\ %
\begin{tikzpicture}
	\begin{pgfonlayer}{nodelayer}
		\node [style=none] (0) at (-2.5, -4.25) {};
		\node [style=none] (1) at (-2.5, -3.25) {};
		\node [style=right label] (2) at (-2.5, -4) {$\A$};
		\node [style=none] (3) at (2.5, -4.25) {};
		\node [style=none] (4) at (2.5, -1.5) {};
		\node [style=right label] (5) at (2.5, -4) {$\B$};
		\node [style=none] (6) at (-1, -3.25) {};
		\node [style=none] (7) at (-1, -4.25) {};
		\node [style=right label] (8) at (-1, -4) {$\G_A$};
		\node [style=none] (9) at (-3, -3.25) {};
		\node [style=none] (10) at (-3, -2.25) {};
		\node [style=none] (11) at (-0.5, -2.25) {};
		\node [style=none] (12) at (-0.5, -3.25) {};
		\node [style=none] (13) at (-2.5, -2.25) {};
		\node [style=none] (14) at (-2.5, 0.25) {};
		\node [style=right label] (15) at (-2.5, -2) {$\A$};
		\node [style=none] (25) at (2.5, 2) {};
		\node [style=none] (26) at (2.5, -0.5) {};
		\node [style=right label] (28) at (2.5, -0.25) {$\B$};
		\node [style=none] (31) at (-2.5, 0.25) {};
		\node [style=none] (75) at (0, -1.5) {};
		\node [style=none] (76) at (0, -4.25) {};
		\node [style=right label] (77) at (0, -4) {$\G_R$};
		\node [style=none] (78) at (1, -1.5) {};
		\node [style=none] (79) at (1, -4.25) {};
		\node [style=right label] (80) at (1, -4) {$\G_B$};
		\node [style=none] (81) at (-1, -1.5) {};
		\node [style=none] (82) at (-1, -2.25) {};
		\node [style=right label] (83) at (-1, -2) {$\G_A$};
		\node [style=none] (92) at (-1, 0.25) {};
		\node [style=none] (93) at (-1, -0.5) {};
		\node [style=right label] (94) at (-1, -0.25) {$\G_A$};
		\node [style=none] (99) at (0, 2) {};
		\node [style=none] (100) at (0, -0.5) {};
		\node [style=right label] (101) at (0, -0.25) {$\G_R$};
		\node [style=none] (102) at (1, 2) {};
		\node [style=none] (103) at (1, -0.5) {};
		\node [style=right label] (104) at (1, -0.25) {$\G_B$};
		\node [style=none] (114) at (-3, 0.25) {};
		\node [style=none] (115) at (-3, 1.25) {};
		\node [style=none] (116) at (-0.5, 1.25) {};
		\node [style=none] (117) at (-0.5, 0.25) {};
		\node [style=none] (118) at (-2.5, 1.25) {};
		\node [style=none] (119) at (-2.5, 3.75) {};
		\node [style=right label] (120) at (-2.5, 1.5) {$\A$};
		\node [style=none] (125) at (2.5, 3.75) {};
		\node [style=none] (126) at (2.5, 3) {};
		\node [style=right label] (127) at (2.5, 3.25) {$\B$};
		\node [style=none] (128) at (-2.5, 3.75) {};
		\node [style=none] (136) at (-1, 2) {};
		\node [style=none] (137) at (-1, 1.25) {};
		\node [style=right label] (138) at (-1, 1.5) {$\G_A$};
		\node [style=none] (146) at (-1, 3.75) {};
		\node [style=none] (147) at (-1, 3) {};
		\node [style=right label] (148) at (-1, 3.25) {$\G_A$};
		\node [style=none] (149) at (0, 3.75) {};
		\node [style=none] (150) at (0, 3) {};
		\node [style=right label] (151) at (0, 3.25) {$\G_R$};
		\node [style=none] (152) at (1, 3.75) {};
		\node [style=none] (153) at (1, 3) {};
		\node [style=right label] (154) at (1, 3.25) {$\G_B$};
		\node [style=none] (155) at (-1.75, 3) {};
		\node [style=none] (156) at (-1.75, 2) {};
		\node [style=none] (157) at (0.25, 2) {};
		\node [style=none] (158) at (0.25, 2) {};
		\node [style=none] (159) at (3.25, 2) {};
		\node [style=none] (160) at (3.25, 3) {};
		\node [style=none] (161) at (1.75, 3) {};
		\node [style=none] (162) at (1.75, 3) {};
		\node [style=none] (163) at (-1.75, -0.5) {};
		\node [style=none] (164) at (-1.75, -1.5) {};
		\node [style=none] (165) at (0.25, -1.5) {};
		\node [style=none] (166) at (0.25, -1.5) {};
		\node [style=none] (167) at (3.25, -1.5) {};
		\node [style=none] (168) at (3.25, -0.5) {};
		\node [style=none] (169) at (1.75, -0.5) {};
		\node [style=none] (170) at (1.75, -0.5) {};
		\node [style=none] (171) at (-3.25, 1.5) {};
		\node [style=none] (172) at (1.5, 1.5) {};
		\node [style=none] (173) at (1.5, 0) {};
		\node [style=none] (174) at (-3.25, 0) {};
		\node [style=none] (175) at (-3.25, -2) {};
		\node [style=none] (176) at (1.5, -2) {};
		\node [style=none] (177) at (1.5, -3.5) {};
		\node [style=none] (178) at (-3.25, -3.5) {};
		\node [style=none] (179) at (-1.25, -3.5) {};
		\node [style=none] (180) at (1.25, -3.5) {};
		\node [style=none] (181) at (1.25, -4.25) {};
		\node [style=none] (182) at (-1.25, -4.25) {};
		\node [style=none] (183) at (-1.25, 3.75) {};
		\node [style=none] (184) at (1.25, 3.75) {};
		\node [style=none] (185) at (1.25, 3) {};
		\node [style=none] (186) at (-1.25, 3) {};
		\node [style=none] (187) at (-1.25, -1.5) {};
		\node [style=none] (188) at (1.25, -1.5) {};
		\node [style=none] (189) at (1.25, -2) {};
		\node [style=none] (190) at (-1.25, -2) {};
		\node [style=none] (191) at (-1.25, 0) {};
		\node [style=none] (192) at (1.25, 0) {};
		\node [style=none] (193) at (1.25, -0.5) {};
		\node [style=none] (194) at (-1.25, -0.5) {};
		\node [style=none] (195) at (-1.25, 2) {};
		\node [style=none] (196) at (1.25, 2) {};
		\node [style=none] (197) at (1.25, 1.5) {};
		\node [style=none] (198) at (-1.25, 1.5) {};
	\end{pgfonlayer}
	\begin{pgfonlayer}{edgelayer}
		\draw [qWire] (1.center) to (0.center);
		\draw [qWire] (4.center) to (3.center);
		\draw (6.center) to (7.center);
		\draw (10.center) to (11.center);
		\draw (11.center) to (12.center);
		\draw (12.center) to (9.center);
		\draw (9.center) to (10.center);
		\draw [qWire] (14.center) to (13.center);
		\draw [qWire] (25.center) to (26.center);
		\draw (75.center) to (76.center);
		\draw (78.center) to (79.center);
		\draw (81.center) to (82.center);
		\draw (92.center) to (93.center);
		\draw (99.center) to (100.center);
		\draw (102.center) to (103.center);
		\draw (115.center) to (116.center);
		\draw (116.center) to (117.center);
		\draw (117.center) to (114.center);
		\draw (114.center) to (115.center);
		\draw [qWire] (119.center) to (118.center);
		\draw [qWire] (125.center) to (126.center);
		\draw (136.center) to (137.center);
		\draw (146.center) to (147.center);
		\draw (149.center) to (150.center);
		\draw (152.center) to (153.center);
		\draw (160.center)
			 to (159.center)
			 to (158.center)
			 to (157.center)
			 to (156.center)
			 to (155.center)
			 to (162.center)
			 to (161.center)
			 to cycle;
		\draw (168.center)
			 to (167.center)
			 to (166.center)
			 to (165.center)
			 to (164.center)
			 to (163.center)
			 to (170.center)
			 to (169.center)
			 to cycle;
		\draw [thick gray dashed edge] (172.center)
			 to (171.center)
			 to (174.center)
			 to (173.center)
			 to cycle;
		\draw [thick gray dashed edge] (176.center)
			 to (175.center)
			 to (178.center)
			 to (177.center)
			 to cycle;
		\draw [thick gray dashed edge] (180.center)
			 to (179.center)
			 to (182.center)
			 to (181.center)
			 to cycle;
		\draw [thick gray dashed edge] (184.center) to (183.center);
		\draw [thick gray dashed edge] (183.center) to (186.center);
		\draw [thick gray dashed edge] (186.center) to (185.center);
		\draw [thick gray dashed edge] (185.center) to (184.center);
		\draw [thick gray dashed edge] (188.center) to (187.center);
		\draw [thick gray dashed edge] (187.center) to (190.center);
		\draw [thick gray dashed edge] (190.center) to (189.center);
		\draw [thick gray dashed edge] (189.center) to (188.center);
		\draw [thick gray dashed edge] (192.center) to (191.center);
		\draw [thick gray dashed edge] (191.center) to (194.center);
		\draw [thick gray dashed edge] (194.center) to (193.center);
		\draw [thick gray dashed edge] (193.center) to (192.center);
		\draw [thick gray dashed edge] (196.center) to (195.center);
		\draw [thick gray dashed edge] (195.center) to (198.center);
		\draw [thick gray dashed edge] (198.center) to (197.center);
		\draw [thick gray dashed edge] (197.center) to (196.center);
	\end{pgfonlayer}
\end{tikzpicture}}\\
&=\ %
\begin{tikzpicture}
	\begin{pgfonlayer}{nodelayer}
		\node [style=none] (0) at (-2, -3.75) {};
		\node [style=none] (1) at (-2, -3) {};
		\node [style=right label] (2) at (-2, -3.5) {$\A$};
		\node [style=none] (3) at (1, -3.75) {};
		\node [style=none] (4) at (1, -1.25) {};
		\node [style=right label] (5) at (1, -3.5) {$\B$};
		\node [style=none] (6) at (-0.5, -3) {};
		\node [style=none] (7) at (-0.5, -3.75) {};
		\node [style=right label] (8) at (-0.5, -3.5) {$\G$};
		\node [style=none] (9) at (-2.5, -3) {};
		\node [style=none] (10) at (-2.5, -2) {};
		\node [style=none] (11) at (0, -2) {};
		\node [style=none] (12) at (0, -3) {};
		\node [style=none] (13) at (-2, -2) {};
		\node [style=none] (14) at (-2, 0.5) {};
		\node [style=right label] (15) at (-2, -1.75) {$\A$};
		\node [style=none] (25) at (1, 2.25) {};
		\node [style=none] (26) at (1, -0.25) {};
		\node [style=right label] (28) at (1, 0) {$\B$};
		\node [style=none] (31) at (-2, 0.5) {};
		\node [style=none] (81) at (-0.5, -1.25) {};
		\node [style=none] (82) at (-0.5, -2) {};
		\node [style=right label] (83) at (-0.5, -1.75) {$\G$};
		\node [style=none] (92) at (-0.5, 0.5) {};
		\node [style=none] (93) at (-0.5, -0.25) {};
		\node [style=right label] (94) at (-0.5, 0) {$\G$};
		\node [style=none] (114) at (-2.5, 0.5) {};
		\node [style=none] (115) at (-2.5, 1.5) {};
		\node [style=none] (116) at (0, 1.5) {};
		\node [style=none] (117) at (0, 0.5) {};
		\node [style=none] (118) at (-2, 1.5) {};
		\node [style=none] (119) at (-2, 4) {};
		\node [style=right label] (120) at (-2, 1.75) {$\A$};
		\node [style=none] (125) at (1, 4) {};
		\node [style=none] (126) at (1, 3.25) {};
		\node [style=right label] (127) at (1, 3.5) {$\B$};
		\node [style=none] (128) at (-2, 4) {};
		\node [style=none] (129) at (1, 4) {};
		\node [style=none] (136) at (-0.5, 2.25) {};
		\node [style=none] (137) at (-0.5, 1.5) {};
		\node [style=right label] (138) at (-0.5, 1.75) {$\G$};
		\node [style=none] (146) at (-0.5, 4) {};
		\node [style=none] (147) at (-0.5, 3.25) {};
		\node [style=right label] (148) at (-0.5, 3.5) {$\G$};
		\node [style=none] (154) at (-1, 2.25) {};
		\node [style=none] (155) at (-1, 3.25) {};
		\node [style=none] (156) at (1.5, 3.25) {};
		\node [style=none] (157) at (1.5, 2.25) {};
		\node [style=none] (158) at (-1, -1.25) {};
		\node [style=none] (159) at (-1, -0.25) {};
		\node [style=none] (160) at (1.5, -0.25) {};
		\node [style=none] (161) at (1.5, -1.25) {};
	\end{pgfonlayer}
	\begin{pgfonlayer}{edgelayer}
		\draw [qWire] (1.center) to (0.center);
		\draw [qWire] (4.center) to (3.center);
		\draw (6.center) to (7.center);
		\draw (10.center) to (11.center);
		\draw (11.center) to (12.center);
		\draw (12.center) to (9.center);
		\draw (9.center) to (10.center);
		\draw [qWire] (14.center) to (13.center);
		\draw [qWire] (25.center) to (26.center);
		\draw (81.center) to (82.center);
		\draw (92.center) to (93.center);
		\draw (115.center) to (116.center);
		\draw (116.center) to (117.center);
		\draw (117.center) to (114.center);
		\draw (114.center) to (115.center);
		\draw [qWire] (119.center) to (118.center);
		\draw [qWire] (125.center) to (126.center);
		\draw (136.center) to (137.center);
		\draw (146.center) to (147.center);
		\draw (155.center) to (156.center);
		\draw (156.center) to (157.center);
		\draw (157.center) to (154.center);
		\draw (154.center) to (155.center);
		\draw (159.center) to (160.center);
		\draw (160.center) to (161.center);
		\draw (161.center) to (158.center);
		\draw (158.center) to (159.center);
	\end{pgfonlayer}
\end{tikzpicture}}.
\end{align}
This can easily be identified as a $2$ round LOCC protocol. Clearly, if we had more layers in the above interaction this generalises to an $n$-round LOCC protocol, i.e.:
\beq
\begin{tikzpicture}
	\begin{pgfonlayer}{nodelayer}
		\node [style=none] (0) at (-2, -7) {};
		\node [style=none] (1) at (-2, -5.25) {};
		\node [style=right label] (2) at (-2, -6.75) {$\A$};
		\node [style=none] (3) at (2, -7) {};
		\node [style=none] (4) at (2, -3.25) {};
		\node [style=right label] (5) at (2, -6.75) {$\B$};
		\node [style=none] (6) at (0, -5.25) {};
		\node [style=point] (7) at (0, -6.25) {$s$};
		\node [style=right label] (8) at (0, -5.75) {$\G$};
		\node [style=none] (9) at (-2.5, -5.25) {};
		\node [style=none] (10) at (-2.5, -4.25) {};
		\node [style=none] (11) at (0.5, -4.25) {};
		\node [style=none] (12) at (0.5, -5.25) {};
		\node [style=none] (13) at (-2, -4.25) {};
		\node [style=none] (14) at (-2, -1.25) {};
		\node [style=right label] (15) at (-2, -1.75) {$\A$};
		\node [style=none] (16) at (0, -3.25) {};
		\node [style=none] (17) at (0, -4.25) {};
		\node [style=right label] (18) at (0, -4) {$\G$};
		\node [style=none] (19) at (-0.5, -3.25) {};
		\node [style=none] (20) at (-0.5, -2.25) {};
		\node [style=none] (21) at (2.5, -2.25) {};
		\node [style=none] (22) at (2.5, -3.25) {};
		\node [style=none] (23) at (0, -2.25) {};
		\node [style=none] (24) at (0, -1.25) {};
		\node [style=none] (25) at (2, -1.25) {};
		\node [style=none] (26) at (2, -2.25) {};
		\node [style=right label] (27) at (0, -1.75) {$\G$};
		\node [style=right label] (28) at (2, -1.75) {$\B$};
		\node [style=none] (29) at (-1, -4.75) {$\mathcal{I}_A^{(1)}$};
		\node [style=none] (30) at (1, -2.75) {$\mathcal{I}_B^{(1)}$};
		\node [style=none] (31) at (-2, -1.25) {};
		\node [style=none] (32) at (2, -1.25) {};
		\node [style=none] (33) at (0, -1.25) {};
		\node [style=none] (56) at (0, 0) {$\vdots$};
		\node [style=none] (57) at (2, 0) {$\vdots$};
		\node [style=none] (58) at (-2, 0) {$\vdots$};
		\node [style=none] (59) at (-1, 0) {$\vdots$};
		\node [style=none] (60) at (1, 0) {$\vdots$};
		\node [style=none] (61) at (-2, 2.25) {};
		\node [style=right label] (62) at (-2, 1.75) {$\A$};
		\node [style=none] (63) at (0, 1.25) {};
		\node [style=none] (64) at (0, 2.25) {};
		\node [style=none] (65) at (2, 4.25) {};
		\node [style=none] (66) at (2, 1.25) {};
		\node [style=right label] (67) at (0, 1.75) {$\G$};
		\node [style=right label] (68) at (2, 1.75) {$\B$};
		\node [style=none] (69) at (-2, 2.25) {};
		\node [style=none] (70) at (2, 4.25) {};
		\node [style=none] (71) at (0, 2.25) {};
		\node [style=none] (72) at (-2.5, 2.25) {};
		\node [style=none] (73) at (-2.5, 3.25) {};
		\node [style=none] (74) at (0.5, 3.25) {};
		\node [style=none] (75) at (0.5, 2.25) {};
		\node [style=none] (76) at (-2, 3.25) {};
		\node [style=none] (77) at (-2, 6.25) {};
		\node [style=right label] (78) at (-2, 5.75) {$\A$};
		\node [style=none] (79) at (0, 4.25) {};
		\node [style=none] (80) at (0, 3.25) {};
		\node [style=right label] (81) at (0, 3.5) {$\G$};
		\node [style=none] (82) at (-0.5, 4.25) {};
		\node [style=none] (83) at (-0.5, 5.25) {};
		\node [style=none] (84) at (2.5, 5.25) {};
		\node [style=none] (85) at (2.5, 4.25) {};
		\node [style=none] (86) at (0, 5.25) {};
		\node [style=none] (87) at (0, 6.25) {};
		\node [style=none] (88) at (2, 6.25) {};
		\node [style=none] (89) at (2, 5.25) {};
		\node [style=right label] (90) at (0, 5.75) {$\G$};
		\node [style=right label] (91) at (2, 5.75) {$\B$};
		\node [style=none] (92) at (-1, 2.75) {$\mathcal{I}_A^{(n)}$};
		\node [style=none] (93) at (1, 4.75) {$\mathcal{I}_B^{(n)}$};
		\node [style=none] (94) at (-2, 1.25) {};
		\node [style=none] (95) at (-2, 2.25) {};
		\node [style=none] (96) at (-2, 2.25) {};
	\end{pgfonlayer}
	\begin{pgfonlayer}{edgelayer}
		\draw [qWire] (1.center) to (0.center);
		\draw [qWire] (4.center) to (3.center);
		\draw (6.center) to (7);
		\draw (10.center) to (11.center);
		\draw (11.center) to (12.center);
		\draw (12.center) to (9.center);
		\draw (9.center) to (10.center);
		\draw [qWire] (14.center) to (13.center);
		\draw (16.center) to (17.center);
		\draw (20.center) to (21.center);
		\draw (21.center) to (22.center);
		\draw (22.center) to (19.center);
		\draw (19.center) to (20.center);
		\draw (24.center) to (23.center);
		\draw [qWire] (25.center) to (26.center);
		\draw (64.center) to (63.center);
		\draw [qWire] (65.center) to (66.center);
		\draw (73.center) to (74.center);
		\draw (74.center) to (75.center);
		\draw (75.center) to (72.center);
		\draw (72.center) to (73.center);
		\draw [qWire] (77.center) to (76.center);
		\draw (79.center) to (80.center);
		\draw (83.center) to (84.center);
		\draw (84.center) to (85.center);
		\draw (85.center) to (82.center);
		\draw (82.center) to (83.center);
		\draw (87.center) to (86.center);
		\draw [qWire] (88.center) to (89.center);
		\draw (95.center) to (94.center);
	\end{pgfonlayer}
\end{tikzpicture}\quad=\quad 
\begin{tikzpicture}
	\begin{pgfonlayer}{nodelayer}
		\node [style=none] (0) at (1.25, -6.75) {};
		\node [style=none] (1) at (1.25, -5) {};
		\node [style=right label] (2) at (1.25, -6.5) {$\A$};
		\node [style=none] (3) at (8, -6.75) {};
		\node [style=none] (4) at (8, -3) {};
		\node [style=right label] (5) at (8, -6.5) {$\B$};
		\node [style=none] (6) at (3.25, -5) {};
		\node [style=point] (7) at (3.25, -6) {$s$};
		\node [style=right label] (8) at (3.25, -5.5) {$\G$};
		\node [style=none] (9) at (0.75, -5) {};
		\node [style=none] (10) at (0.75, -4) {};
		\node [style=none] (11) at (3.75, -4) {};
		\node [style=none] (12) at (3.75, -5) {};
		\node [style=none] (13) at (1.25, -4) {};
		\node [style=none] (14) at (1.25, -1) {};
		\node [style=right label] (15) at (1.25, -1.5) {$\A$};
		\node [style=none] (16) at (6, -3) {};
		\node [style=none] (17) at (3.25, -4) {};
		\node [style=right label] (18) at (4.75, -3.75) {$\G$};
		\node [style=none] (19) at (5.5, -3) {};
		\node [style=none] (20) at (5.5, -2) {};
		\node [style=none] (21) at (8.5, -2) {};
		\node [style=none] (22) at (8.5, -3) {};
		\node [style=none] (23) at (6, -2) {};
		\node [style=none] (24) at (3.75, -1) {};
		\node [style=none] (25) at (8, -1) {};
		\node [style=none] (26) at (8, -2) {};
		\node [style=right label] (27) at (4.75, -1.75) {$\G$};
		\node [style=right label] (28) at (8, -1.5) {$\B$};
		\node [style=none] (29) at (2.25, -4.5) {$\mathcal{I}_A^{(1)}$};
		\node [style=none] (30) at (7, -2.5) {$\mathcal{I}_B^{(1)}$};
		\node [style=none] (31) at (1.25, -1) {};
		\node [style=none] (32) at (8, -1) {};
		\node [style=none] (56) at (4, 0.25) {$\vdots$};
		\node [style=none] (57) at (8, 0.25) {$\vdots$};
		\node [style=none] (58) at (1.25, 0.25) {$\vdots$};
		\node [style=none] (59) at (2.25, 0.25) {$\vdots$};
		\node [style=none] (60) at (7, 0.25) {$\vdots$};
		\node [style=none] (61) at (1.25, 2.5) {};
		\node [style=right label] (62) at (1.25, 2) {$\A$};
		\node [style=none] (63) at (5.75, 1.5) {};
		\node [style=none] (64) at (3.5, 2.5) {};
		\node [style=none] (65) at (8, 4.5) {};
		\node [style=none] (66) at (8, 1.5) {};
		\node [style=right label] (67) at (4.75, 2) {$\G$};
		\node [style=right label] (68) at (8, 2) {$\B$};
		\node [style=none] (69) at (8, 4.5) {};
		\node [style=none] (70) at (3.5, 2.5) {};
		\node [style=none] (71) at (0.75, 2.5) {};
		\node [style=none] (72) at (0.75, 3.5) {};
		\node [style=none] (73) at (4, 3.5) {};
		\node [style=none] (74) at (4, 2.5) {};
		\node [style=none] (75) at (1.25, 3.5) {};
		\node [style=none] (76) at (1.25, 6.5) {};
		\node [style=right label] (77) at (1.25, 6) {$\A$};
		\node [style=none] (78) at (6, 4.5) {};
		\node [style=none] (79) at (3.5, 3.5) {};
		\node [style=right label] (80) at (4.75, 3.75) {$\G$};
		\node [style=none] (81) at (5.5, 4.5) {};
		\node [style=none] (82) at (5.5, 5.5) {};
		\node [style=none] (83) at (8.5, 5.5) {};
		\node [style=none] (84) at (8.5, 4.5) {};
		\node [style=none] (85) at (6, 5.5) {};
		\node [style=none] (86) at (6, 6.5) {};
		\node [style=none] (87) at (8, 6.5) {};
		\node [style=none] (88) at (8, 5.5) {};
		\node [style=right label] (89) at (6, 6) {$\G$};
		\node [style=right label] (90) at (8, 6) {$\B$};
		\node [style=none] (91) at (2.25, 3) {$\mathcal{I}_A^{(n)}$};
		\node [style=none] (92) at (7, 5) {$\mathcal{I}_B^{(n)}$};
		\node [style=none] (93) at (4.75, 7.25) {};
		\node [style=none] (94) at (4.75, -8) {};
		\node [style=none] (95) at (5.75, 0.25) {$\vdots$};
		\node [style=none] (96) at (2.5, -7.75) {\color{gray}Alice};
		\node [style=none] (97) at (6.25, -7.75) {\color{gray}Bob};
		\node [style=none] (98) at (1.25, 2.5) {};
		\node [style=none] (99) at (1.25, 1.5) {};
	\end{pgfonlayer}
	\begin{pgfonlayer}{edgelayer}
		\draw [qWire] (1.center) to (0.center);
		\draw [qWire] (4.center) to (3.center);
		\draw (6.center) to (7);
		\draw (10.center) to (11.center);
		\draw (11.center) to (12.center);
		\draw (12.center) to (9.center);
		\draw (9.center) to (10.center);
		\draw [qWire] (14.center) to (13.center);
		\draw [in=90, out=-90, looseness=0.50] (16.center) to (17.center);
		\draw (20.center) to (21.center);
		\draw (21.center) to (22.center);
		\draw (22.center) to (19.center);
		\draw (19.center) to (20.center);
		\draw [in=90, out=-90, looseness=0.50] (24.center) to (23.center);
		\draw [qWire] (25.center) to (26.center);
		\draw [in=150, out=-90] (64.center) to (63.center);
		\draw [qWire] (65.center) to (66.center);
		\draw (72.center) to (73.center);
		\draw (73.center) to (74.center);
		\draw (74.center) to (71.center);
		\draw (71.center) to (72.center);
		\draw [qWire] (76.center) to (75.center);
		\draw [in=90, out=-90, looseness=0.50] (78.center) to (79.center);
		\draw (82.center) to (83.center);
		\draw (83.center) to (84.center);
		\draw (84.center) to (81.center);
		\draw (81.center) to (82.center);
		\draw (86.center) to (85.center);
		\draw [qWire] (87.center) to (88.center);
		\draw [thick gray dashed edge, in=90, out=-90] (93.center) to (94.center);
		\draw [qWire] (98.center) to (99.center);
	\end{pgfonlayer}
\end{tikzpicture}.
\eeq
It is well known that, even in GPTs, entanglement cannot be generated by LOCC.

To see this explicitly, let us consider inserting resolutions of the identity and postselecting on a final state of the field\footnote{Note that typically we would consider the trivial postselection, i.e., simply ignoring the final state of the field as it is not directly accessible, here we allow for arbitrary postselection purely for the sake of generality.}:
\begin{align}
\begin{tikzpicture}
	\begin{pgfonlayer}{nodelayer}
		\node [style=none] (0) at (2, -11.75) {};
		\node [style=none] (1) at (2, -10) {};
		\node [style=right label] (2) at (2, -11.5) {$\A$};
		\node [style=none] (3) at (6, -11.5) {};
		\node [style=none] (4) at (6, -5.5) {};
		\node [style=right label] (5) at (6, -11.25) {$\B$};
		\node [style=none] (6) at (4, -10) {};
		\node [style=point] (7) at (4, -11) {$s$};
		\node [style=right label] (8) at (4, -10.5) {$\G$};
		\node [style=none] (9) at (1.5, -10) {};
		\node [style=none] (10) at (1.5, -9) {};
		\node [style=none] (11) at (4.5, -9) {};
		\node [style=none] (12) at (4.5, -10) {};
		\node [style=none] (13) at (2, -9) {};
		\node [style=none] (14) at (2, -1) {};
		\node [style=right label] (15) at (2, -1.5) {$\A$};
		\node [style=none] (16) at (3.5, -5.5) {};
		\node [style=none] (17) at (3.5, -4.5) {};
		\node [style=none] (18) at (6.5, -4.5) {};
		\node [style=none] (19) at (6.5, -5.5) {};
		\node [style=none] (20) at (6, -1) {};
		\node [style=none] (21) at (6, -4.5) {};
		\node [style=right label] (22) at (6, -1.5) {$\B$};
		\node [style=none] (23) at (3, -9.5) {$\mathcal{I}_A^{(1)}$};
		\node [style=none] (24) at (5, -5) {$\mathcal{I}_B^{(1)}$};
		\node [style=none] (25) at (2, -1) {};
		\node [style=none] (26) at (4.75, 0) {$\vdots$};
		\node [style=none] (27) at (6, 0) {$\vdots$};
		\node [style=none] (28) at (2, 0) {$\vdots$};
		\node [style=none] (29) at (3.25, 0) {$\vdots$};
		\node [style=none] (30) at (4.75, 0) {$\vdots$};
		\node [style=none] (31) at (2, 4.5) {};
		\node [style=right label] (32) at (2, 1.25) {$\A$};
		\node [style=none] (33) at (6, 9) {};
		\node [style=none] (34) at (6, 1) {};
		\node [style=right label] (35) at (6, 1.25) {$\B$};
		\node [style=none] (36) at (2, 4.5) {};
		\node [style=none] (37) at (6, 9) {};
		\node [style=none] (38) at (1.5, 4.5) {};
		\node [style=none] (39) at (1.5, 5.5) {};
		\node [style=none] (40) at (4.5, 5.5) {};
		\node [style=none] (41) at (4.5, 4.5) {};
		\node [style=none] (42) at (2, 5.5) {};
		\node [style=none] (43) at (2, 11.5) {};
		\node [style=right label] (44) at (2, 11) {$\A$};
		\node [style=none] (45) at (3.5, 9) {};
		\node [style=none] (46) at (3.5, 10) {};
		\node [style=none] (47) at (6.5, 10) {};
		\node [style=none] (48) at (6.5, 9) {};
		\node [style=none] (49) at (6, 11.5) {};
		\node [style=none] (50) at (6, 10) {};
		\node [style=right label] (51) at (6, 11) {$\B$};
		\node [style=none] (52) at (3, 5) {$\mathcal{I}_A^{(n)}$};
		\node [style=none] (53) at (5, 9.5) {$\mathcal{I}_B^{(n)}$};
		\node [style=none] (54) at (2, 1) {};
		\node [style=none] (55) at (2, 4.5) {};
		\node [style=none] (56) at (2, 4.5) {};
		\node [style=copoint, fill=black] (57) at (4, -8) {$\color{white}a_1$};
		\node [style=none] (58) at (4, -9) {};
		\node [style=right label] (59) at (4, -8.75) {$\G$};
		\node [style=none] (60) at (4, 10) {};
		\node [style=copoint, fill=black] (61) at (4, 11) {$\color{white}s'$};
		\node [style=right label] (62) at (4, 10.25) {$\G$};
		\node [style=point] (63) at (4, -6.5) {$a_1$};
		\node [style=none] (64) at (4, -5.5) {};
		\node [style=right label] (65) at (4, -6) {$\G$};
		\node [style=none] (66) at (3, -7.25) {$\displaystyle\sum_{a_1}$};
		\node [style=copoint, fill=black] (67) at (4, -3.5) {$\color{white}a_2$};
		\node [style=none] (68) at (4, -4.5) {};
		\node [style=right label] (69) at (4, -4.25) {$\G$};
		\node [style=point] (70) at (4, -2.25) {$a_2$};
		\node [style=none] (71) at (4, -1) {};
		\node [style=right label] (72) at (4, -1.5) {$\G$};
		\node [style=none] (73) at (3, -3) {$\displaystyle\sum_{a_2}$};
		\node [style=copoint, fill=black] (74) at (4, 2) {$\color{white}n_1$};
		\node [style=none] (75) at (4, 1) {};
		\node [style=right label] (76) at (4, 1.25) {$\G$};
		\node [style=point] (77) at (4, 3.5) {$n_1$};
		\node [style=none] (78) at (4, 4.5) {};
		\node [style=right label] (79) at (4, 4) {$\G$};
		\node [style=none] (80) at (3, 2.75) {$\displaystyle\sum_{n_1}$};
		\node [style=copoint, fill=black] (81) at (4, 6.5) {$\color{white}n_2$};
		\node [style=none] (82) at (4, 5.5) {};
		\node [style=right label] (83) at (4, 5.75) {$\G$};
		\node [style=point] (84) at (4, 8) {$n_2$};
		\node [style=none] (85) at (4, 9) {};
		\node [style=right label] (86) at (4, 8.5) {$\G$};
		\node [style=none] (87) at (3, 7.25) {$\displaystyle\sum_{n_2}$};
	\end{pgfonlayer}
	\begin{pgfonlayer}{edgelayer}
		\draw [qWire] (1.center) to (0.center);
		\draw [qWire] (4.center) to (3.center);
		\draw (6.center) to (7);
		\draw (10.center) to (11.center);
		\draw (11.center) to (12.center);
		\draw (12.center) to (9.center);
		\draw (9.center) to (10.center);
		\draw [qWire] (14.center) to (13.center);
		\draw (17.center) to (18.center);
		\draw (18.center) to (19.center);
		\draw (19.center) to (16.center);
		\draw (16.center) to (17.center);
		\draw [qWire] (20.center) to (21.center);
		\draw [qWire] (33.center) to (34.center);
		\draw (39.center) to (40.center);
		\draw (40.center) to (41.center);
		\draw (41.center) to (38.center);
		\draw (38.center) to (39.center);
		\draw [qWire] (43.center) to (42.center);
		\draw (46.center) to (47.center);
		\draw (47.center) to (48.center);
		\draw (48.center) to (45.center);
		\draw (45.center) to (46.center);
		\draw [qWire] (49.center) to (50.center);
		\draw (55.center) to (54.center);
		\draw (57) to (58.center);
		\draw (61) to (60.center);
		\draw (63) to (64.center);
		\draw (67) to (68.center);
		\draw (70) to (71.center);
		\draw (74) to (75.center);
		\draw (77) to (78.center);
		\draw (81) to (82.center);
		\draw (84) to (85.center);
	\end{pgfonlayer}
\end{tikzpicture}
\ \ &= \ \ \begin{tikzpicture}
	\begin{pgfonlayer}{nodelayer}
		\node [style=none] (0) at (-3.5, -5.75) {};
		\node [style=none] (1) at (-3.5, -3) {};
		\node [style=right label] (2) at (-3.5, -5.5) {$\A$};
		\node [style=none] (3) at (2.25, -5.75) {};
		\node [style=none] (4) at (2.25, -3) {};
		\node [style=right label] (5) at (2.25, -5.5) {$\B$};
		\node [style=none] (6) at (-1.5, -3) {};
		\node [style=point] (7) at (-1.5, -4) {$s$};
		\node [style=right label] (8) at (-1.5, -3.5) {$\G$};
		\node [style=none] (9) at (-4, -3) {};
		\node [style=none] (10) at (-4, -2) {};
		\node [style=none] (11) at (-1, -2) {};
		\node [style=none] (12) at (-1, -3) {};
		\node [style=none] (13) at (-3.5, -2) {};
		\node [style=none] (14) at (-3.5, -1) {};
		\node [style=right label] (15) at (-3.5, -1.5) {$\A$};
		\node [style=none] (19) at (-0.25, -3) {};
		\node [style=none] (20) at (-0.25, -2) {};
		\node [style=none] (21) at (2.75, -2) {};
		\node [style=none] (22) at (2.75, -3) {};
		\node [style=none] (25) at (2.25, -1) {};
		\node [style=none] (26) at (2.25, -2) {};
		\node [style=right label] (28) at (2.25, -1.5) {$\B$};
		\node [style=none] (29) at (-2.5, -2.5) {$\mathcal{I}_A^{(1)}$};
		\node [style=none] (30) at (1.25, -2.5) {$\mathcal{I}_B^{(1)}$};
		\node [style=none] (31) at (-3.5, -1) {};
		\node [style=none] (57) at (2.25, 0) {$\vdots$};
		\node [style=none] (58) at (-3.5, 0) {$\vdots$};
		\node [style=none] (59) at (-2.25, 0) {$\vdots$};
		\node [style=none] (60) at (1, 0) {$\vdots$};
		\node [style=none] (61) at (-3.5, 2) {};
		\node [style=right label] (62) at (-3.5, 1.25) {$\A$};
		\node [style=none] (65) at (2.25, 2) {};
		\node [style=none] (66) at (2.25, 1) {};
		\node [style=right label] (68) at (2.25, 1.25) {$\B$};
		\node [style=none] (69) at (-3.5, 2) {};
		\node [style=none] (70) at (2.25, 2) {};
		\node [style=none] (72) at (-4, 2) {};
		\node [style=none] (73) at (-4, 3) {};
		\node [style=none] (74) at (-1, 3) {};
		\node [style=none] (75) at (-1, 2) {};
		\node [style=none] (76) at (-3.5, 3) {};
		\node [style=none] (77) at (-3.5, 6) {};
		\node [style=right label] (78) at (-3.5, 5.75) {$\A$};
		\node [style=none] (82) at (-0.25, 2) {};
		\node [style=none] (83) at (-0.25, 3) {};
		\node [style=none] (84) at (2.75, 3) {};
		\node [style=none] (85) at (2.75, 2) {};
		\node [style=none] (88) at (2.25, 6) {};
		\node [style=none] (89) at (2.25, 3) {};
		\node [style=right label] (91) at (2.25, 5.75) {$\B$};
		\node [style=none] (92) at (-2.5, 2.5) {$\mathcal{I}_A^{(n)}$};
		\node [style=none] (93) at (1.25, 2.5) {$\mathcal{I}_B^{(n)}$};
		\node [style=none] (94) at (-3.5, 1) {};
		\node [style=none] (95) at (-3.5, 2) {};
		\node [style=none] (96) at (-3.5, 2) {};
		\node [style=copoint, fill=black] (113) at (-1.5, -1) {$\color{white}a_1$};
		\node [style=none] (114) at (-1.5, -2) {};
		\node [style=right label] (115) at (-1.5, -1.75) {$\G$};
		\node [style=none] (120) at (0.25, 3) {};
		\node [style=copoint, fill=black] (121) at (0.25, 4) {$\color{white}s'$};
		\node [style=right label] (124) at (0.25, 3.25) {$\G$};
		\node [style=point] (128) at (0.25, -4) {$a_1$};
		\node [style=none] (129) at (0.25, -3) {};
		\node [style=right label] (130) at (0.25, -3.5) {$\G$};
		\node [style=copoint, fill=black] (132) at (0.25, -1) {$\color{white}a_2$};
		\node [style=none] (133) at (0.25, -2) {};
		\node [style=right label] (134) at (0.25, -1.75) {$\G$};
		\node [style=point] (142) at (-1.5, 1) {$n_1$};
		\node [style=none] (143) at (-1.5, 2) {};
		\node [style=right label] (144) at (-1.5, 1.5) {$\G$};
		\node [style=copoint, fill=black] (146) at (-1.5, 4) {$\color{white}n_2$};
		\node [style=none] (147) at (-1.5, 3) {};
		\node [style=right label] (148) at (-1.5, 3.25) {$\G$};
		\node [style=point] (149) at (0.25, 1) {$n_2$};
		\node [style=none] (150) at (0.25, 2) {};
		\node [style=right label] (151) at (0.25, 1.5) {$\G$};
		\node [style=none] (152) at (-5.75, 0) {$\displaystyle\sum_{a_1,...,n_2}$};
		\node [style=none] (153) at (-4.25, 5.25) {};
		\node [style=none] (154) at (-0.75, 5.25) {};
		\node [style=none] (155) at (-0.75, -5) {};
		\node [style=none] (156) at (-4.25, -5) {};
		\node [style=none] (157) at (-0.5, 5.25) {};
		\node [style=none] (158) at (3, 5.25) {};
		\node [style=none] (159) at (3, -5) {};
		\node [style=none] (160) at (-0.5, -5) {};
	\end{pgfonlayer}
	\begin{pgfonlayer}{edgelayer}
		\draw [qWire] (1.center) to (0.center);
		\draw [qWire] (4.center) to (3.center);
		\draw (6.center) to (7);
		\draw (10.center) to (11.center);
		\draw (11.center) to (12.center);
		\draw (12.center) to (9.center);
		\draw (9.center) to (10.center);
		\draw [qWire] (14.center) to (13.center);
		\draw (20.center) to (21.center);
		\draw (21.center) to (22.center);
		\draw (22.center) to (19.center);
		\draw (19.center) to (20.center);
		\draw [qWire] (25.center) to (26.center);
		\draw [qWire] (65.center) to (66.center);
		\draw (73.center) to (74.center);
		\draw (74.center) to (75.center);
		\draw (75.center) to (72.center);
		\draw (72.center) to (73.center);
		\draw [qWire] (77.center) to (76.center);
		\draw (83.center) to (84.center);
		\draw (84.center) to (85.center);
		\draw (85.center) to (82.center);
		\draw (82.center) to (83.center);
		\draw [qWire] (88.center) to (89.center);
		\draw (95.center) to (94.center);
		\draw (113) to (114.center);
		\draw (121) to (120.center);
		\draw (128) to (129.center);
		\draw (132) to (133.center);
		\draw (142) to (143.center);
		\draw (146) to (147.center);
		\draw (149) to (150.center);
		\draw [thick gray dashed edge] (153.center)
			 to (156.center)
			 to (155.center)
			 to (154.center)
			 to cycle;
		\draw [thick gray dashed edge] (157.center)
			 to (160.center)
			 to (159.center)
			 to (158.center)
			 to cycle;
	\end{pgfonlayer}
\end{tikzpicture} \\
& = \ \ \begin{tikzpicture}
	\begin{pgfonlayer}{nodelayer}
		\node [style=none] (152) at (-5.75, 0) {$\displaystyle\sum_{a_1,...,n_2}$};
		\node [style=none] (161) at (-3, -2.25) {};
		\node [style=none] (162) at (-3, -0.75) {};
		\node [style=right label] (163) at (-3, -2) {$\A$};
		\node [style=none] (164) at (0, -2.25) {};
		\node [style=none] (165) at (0, -0.75) {};
		\node [style=right label] (166) at (0, -2) {$\B$};
		\node [style=none] (167) at (-4.25, -0.75) {};
		\node [style=none] (168) at (-4.25, 0.75) {};
		\node [style=none] (169) at (-1.75, 0.75) {};
		\node [style=none] (170) at (-1.75, -0.75) {};
		\node [style=none] (171) at (-3, 0.75) {};
		\node [style=none] (172) at (-3, 2.25) {};
		\node [style=right label] (173) at (-3, 1.75) {$\A$};
		\node [style=none] (174) at (-1.25, -0.75) {};
		\node [style=none] (175) at (-1.25, 0.75) {};
		\node [style=none] (176) at (1.25, 0.75) {};
		\node [style=none] (177) at (1.25, -0.75) {};
		\node [style=none] (178) at (0, 2.25) {};
		\node [style=none] (179) at (0, 0.75) {};
		\node [style=right label] (180) at (0, 1.75) {$\B$};
		\node [style=none] (181) at (-3, 0) {$\color{white}\mathcal{E}_A^{a_1,..,n_2}$};
		\node [style=none] (182) at (0, 0) {$\color{white}\mathcal{E}_B^{a_1,..,n_2}$};
	\end{pgfonlayer}
	\begin{pgfonlayer}{edgelayer}
		\draw [qWire] (162.center) to (161.center);
		\draw [qWire] (165.center) to (164.center);
		\draw [fill=black] (170.center)
			 to (167.center)
			 to (168.center)
			 to (169.center)
			 to cycle;
		\draw [qWire] (172.center) to (171.center);
		\draw [fill=black] (177.center)
			 to (174.center)
			 to (175.center)
			 to (176.center)
			 to cycle;
		\draw [qWire] (178.center) to (179.center);
	\end{pgfonlayer}
\end{tikzpicture}
 \ \ .
\end{align}
This is of the same form as \eqref{eq:proof1} and so, as we showed above, this cannot generate entanglement.
\endproof

\section{The Newtonian potential as a non physical degree of freedom}
\label{sec:GravityMed}

\begin{figure}[h]
	\begin{center}
		\includegraphics[scale=0.3]{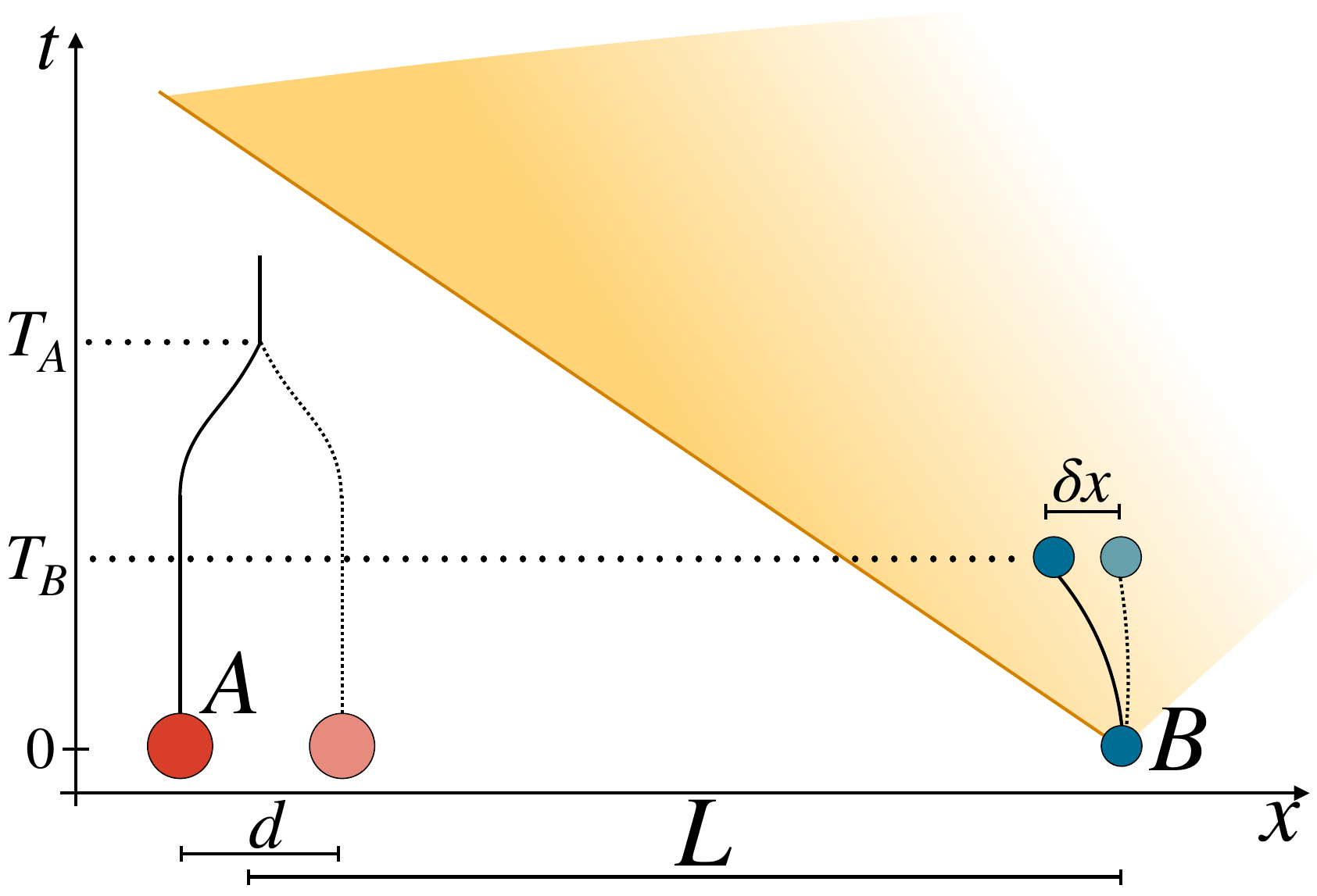}
		\caption{\label{fig:protocol} \textbf{Illustration of the protocol.} Long before the experiment starts, a massive particle $\A$ with mass $m_\A$ is put in a spatial coherent superposition in such a way that the distance of the position of its centre of mass in the two amplitudes is $d$. At time $t=0$ an observer $\B$, located at a distance $L$ from $\A$, can decide whether to release a probe particle $\B$ with mass $m_\B$, initially kept in a trap. If $\B$ does not release the trap, the state of the full system stays separable. If $\B$ releases the trap, and assuming that gravity does not have a (GPT) state associated to it, the state of $\A$'s and $\B$'s particles gets entangled in a time $T_\B$. $\A$ can then recombine her state, ending the experiment at time $T_\A$. With an appropriate choice of the parameters $L,d, T_\A, T_\B, m_\A, m_\B$, and by performing an interference experiment, $\A$ can tell whether $\B$ released the trap or not before a light-crossing time (in the figure, the light cone is illustrated as an orange-shaded area). Hence, if gravity is not treated as a physical degree of freedom, we run into a violation of no-faster-than-light signalling.}
	\end{center}
\end{figure}

In quantum field theory, when deriving the Newtonian limit of a linearized quantum gravity model, the Newtonian potential is fully determined by the matter configuration, and is thus not an independent degree of freedom. This dependence of the Newton potential on the matter configuration has usually been used to argue that the Newtonian potential is not a \emph{physical} mediator. According to this reasoning, experiments only involving the Newtonian potential cannot conclude anything on the nature of the gravitational field. Here, we distinguish the two conditions of \emph{independence} and \emph{physicality}, and show that either we assign a state to the Newtonian potential (physicality), or we run into faster-than-light signalling. This result is a corollary of Refs.~\cite{belenchia2018quantum, belenchia2019information}. 

Requesting that a theory is compatible with the no-faster-than-light principle is a very conservative requirement. In order to show that failing to assign a state to the Newtonian potential leads to a violation of the no-faster-than-light-signalling principle, we review a protocol first introduced in Ref.~\cite{mari2016experiments} in the electromagnetic case, and then studied in Refs.~\cite{belenchia2018quantum, belenchia2019information} in the gravitational case. The protocol is illustrated in Fig.~\ref{fig:protocol}. We consider two observers, $\A$ and $\B$, who are located at distance $L$ from each other. Long before the experiment starts, 1) $\A$ prepares a massive particle of mass $m_\A$ in a spatial coherent superposition state, such that the separation between the centre of mass of $\A$'s particle in the two amplitudes is $d \ll L$, and 2) $\B$ localises a probe particle with mass $m_\B \ll m_\A$ in a trap. At time $t=0$, $\B$ decides whether to release the trap or not.

If the Newtonian potential is not a physical degree of freedom, there is no physical state associated to its left ($L$) or right ($R$) configuration.

Firstly, let us consider the situation in which $\B$ does not release its trap. The full state of $\A$ and $\B$ is in this case $\ket{\Psi} = \frac{1}{\sqrt{2}} ( \ket{L}_\A
+ \ket{R}_\A) \ket{0}_\B$, where $\ket{L}_\A$ and $\ket{R}_\A$ stand for the left and right amplitude of the state of the centre of mass of $\A$'s particle, and $\ket{0}_\B$ stands for the initial position of $\B$'s particle in the trap. This state is unchanged for the whole duration of the experiment.

If $\B$ instead releases the trap, the state of $\B$'s particle starts to become entangled with the state of $\A$'s particle. Calling $T_\B$ the time it takes for the state of $\A$ and $\B$ to become entangled, at $T_\B$ the full state of $\A$ and $\B$ is $\ket{\Phi} = \frac{1}{\sqrt{2}} ( \ket{L}_\A  \ket{L}_\B
+ \ket{R}_\A \ket{R}_\B)$. Similarly to the case when $\B$ does not release the trap, $\A$ can now recombine her superposition, ending her experiment at time $T_\A$. It is possible to find a combination of the parameters $L, d, T_\A, T_\B, m_\A, m_\B$ such that $\A$ can recombine her state before a light-crossing time.

Assuming the standard Born rule, it is now clear that the outcome of $\A$'s experiment is different according to whether $\B$ decided to release the trap or not. If $\B$ did not release the trap, $\A$ sees an interference pattern when recombining the state; if $\B$ releases the trap, the state decoheres and $\A$ does not see any interference. Hence, $\A$ is able to know whether $\B$ released the trap or not before a light-crossing time, thus violating the no-faster-than-light signalling principle. In Refs.~\cite{belenchia2018quantum, belenchia2019information} it was shown that, in a quantum formalism in which gravity is a mediator and can become entangled with the states of $\A$ and $\B$, then there is no violation of faster-than-light signalling. It is relevant to notice that the fact that the Newtonian potential is responsible for the entanglement is no approximation here, and that the violation of the no-faster-than-light signalling principle would be the same if, for instance, we considered a retarded potential. Analogously, no propagating degrees of freedom of gravity are to be considered before $\A$ starts recombining her superposition because, when the experiment starts, particle $\A$ and the gravitational field are already in a static configuration. It was then argued in Ref.~\cite{belenchia2019information} that a mass in a spatial superposition should be considered entangled with its own Newtonian field, and that the Newtonian field can carry (quantum) information.

Observe that the notion of faster-than-light-signalling here is distinct from what is sometimes called `no-signalling' in the GPT literature (which we call subsystem independence). Subsystem independence states that the reduced state of one subsystem is independent of operations on the other subsystem. In the above argument however Bob can choose one of two states $\ket 0$ or $\ket{L} + \ket{R}$. A global unitary is then applied (given by $V(x_1,x_2)$), and depending on Bob's choice of initial state the global evolution leads to Alice's reduced state being either a pure state or a mixed state. The key feature of this argument is that, given a joint evolution $U$ of two systems $\A$ and $\B$, then different input states for $\B$ can give different reduced states for $\A$, and that this change in reduced state for $\A$ occurs before a light signal can be transmitted between parties. This is completely consistent with no-signalling in a GPT sense, that is, with $\A$ and $\B$ being independent subsystems.

The above argument is valid in quantum theory, however assigning a state to the Newtonian field also finds a justification in GPTs, where all elements of the description are treated as systems, on which further constraints are then imposed.

\bibliography{biblio}{}
\bibliographystyle{quantum}

\end{document}